\documentclass[lettersize,journal]{IEEEtran}
\usepackage{cite}
\usepackage{amsmath,amssymb,amsfonts}
\usepackage{algorithmic}
\usepackage{graphicx}
\usepackage{textcomp}
\usepackage{xcolor}
\usepackage{graphicx}
\usepackage{caption}
\graphicspath{{./figure/}}
\usepackage{soul}
\usepackage{color}
\usepackage{multirow}
\usepackage{amsmath}

\usepackage{amssymb}
\usepackage[linesnumbered,ruled]{algorithm2e}
\usepackage{booktabs}
\usepackage{array}
\usepackage{subfig}
\usepackage{url}
\usepackage{tabularx}
\usepackage{bm}
\usepackage{enumitem}
\usepackage{pifont}
\newcommand{\matr}[1]{\bm{#1}}
\newcommand{\vect}[1]{\bm{#1}}
\usepackage{xcolor}
\usepackage{textcomp}
\usepackage[commandnameprefix=ifneeded]{changes}
\usepackage{makecell}
\usepackage{mathrsfs}
\usepackage{listings}
\begin{document}

\title{Precision-Aware Iterative Algorithms Based on Group-Shared Exponents of Floating-Point Numbers}

\author{Jianhua Gao, Jiayuan Shen, Yuxiang Zhang, Weixing Ji\textsuperscript{*}, Hua Huang
\thanks{
\textsuperscript{*}Weixing Ji is the corresponding author.}
\thanks{Jianhua Gao, Yuxiang Zhang, Weixing Ji, and Hua Huang are with the School of Artificial Intelligence, Beijing Normal University, Beijing 100875, China (e-mail: gaojh@bnu.edu.cn, zyx@mail.bnu.edu.cn, jwx@bnu.edu.cn, huahuang@bnu.edu.cn).}
\thanks{Jiayuan Shen is with the School of Computer Science and Technology, Beijing Institute of Technology, Beijing 100081, China (e-mail: 3220231249@bit.edu.cn).}
}

\markboth{Journal of \LaTeX\ Class Files,~Vol.~14, No.~8, August~2021}%
{Shell \MakeLowercase{\textit{et al.}}: A Sample Article Using IEEEtran.cls for IEEE Journals}


\maketitle

\begin{abstract}
Iterative solvers are frequently used in scientific applications and engineering computations. However, the memory-bound Sparse Matrix-Vector (SpMV) kernel computation hinders the efficiency of iterative algorithms. As modern hardware increasingly supports low-precision computation, the mixed-precision optimization of iterative algorithms has garnered widespread attention. Nevertheless, existing mixed-precision methods pose challenges, including format conversion overhead, tight coupling between storage and computation representation, and the need to store multiple precision copies of data. This paper proposes a floating-point representation based on the group-shared exponent and segmented storage of the mantissa, enabling higher bit utilization of the representation vector and fast switches between different precisions without needing multiple data copies. Furthermore, a stepped mixed-precision iterative algorithm is proposed. Our experimental results demonstrate that, compared with existing floating-point formats, our approach significantly improves iterative algorithms' performance and convergence residuals.

\end{abstract}


\section{Introduction}
Numerical simulations of many practical problems often rely on solving the linear equation system $\matr{A}\vect{x} = \vect{b}$, where $\matr{A}$ is an $m \times n$ sparse matrix, and $\vect{x}$ and $\vect{b}$ are dense vectors of sizes $n$ and $m$, respectively. Iterative solvers such as conjugate gradient (CG) and generalized minimal residual (GMRES) methods are commonly used in practice. Sparse matrix-vector multiplication (SpMV) is a frequently used computational kernel in these algorithms, but it is also their primary performance bottleneck. The computational efficiency of SpMV is constrained by the memory access efficiency, making it a typical memory-bound kernel.


With the continuous development of artificial intelligence technology, many computer hardware manufacturers have begun to support multiple precision data representations in their latest products. For example, NVIDIA's latest H100 supports TF32, BF16, FP16, FP8 (E4M3 and E5M2), among others. This has inspired researchers to use mixed-precision techniques to optimize iterative algorithms. Compared to traditional iterative algorithms, which typically use a single floating-point format like FP64 or FP32, mixed-precision optimization introduces multiple-precision floating-point representations in iterative algorithms, reducing memory access and computational overhead, thereby improving the efficiency of iterative solving algorithms. As a key kernel of iterative algorithms, mixed-precision optimization of SpMV has attracted the attention of researchers \cite{10.1145/3371275,DBLP:journals/jcam/MukunokiO20,DBLP:conf/sbac-pad/TezcanTKKU22,graillat:hal-03561193,DBLP:journals/jocs/Isupov22,DBLP:journals/corr/Kouya14a,DBLP:conf/cluster/MukunokiI16,graillat2024reduced,zhao2024block,mukunoki2023sparse}. The main idea behind these efforts is to select different-precision floating-point representations for non-zero elements (non-zeros) based on their magnitudes. For example, smaller values are stored in the single-precision floating-point format, while larger values are stored in the double-precision format, with both computations performed in double precision. In terms of mixed-precision optimization for iterative algorithms, current work mainly focuses on using different-precision floating-point representations in different statements or iterations \cite{DBLP:journals/siamsc/CarsonH18,DBLP:journals/corr/abs-1907-10550,DBLP:journals/corr/abs-2009-12101,le2018mixed,loe2021study}.

Existing mixed-precision algorithms have improved the computational efficiency of iterative algorithms. However, using general-purpose floating-point representations introduces several challenges:  
(1)~\textbf{Lower utilization of binary bits}: FP64 and FP32 have 11 and 8 exponent bits, respectively, and are two widely used floating-point representations. However, in specific applications, floating-point data may have particular numerical distributions, such as a small distribution range or most data being distributed around some points. In this case, the exponent field of the existing floating-point representations has a low binary bits utilization.
(2)~\textbf{Tight coupling between storage and computation}: In systems based on traditional floating-point formats, the precision used for storage is tightly coupled with the precision for computation, which sacrifices flexibility and limits the mixed-precision optimization of iterative algorithms.

\begin{figure*}
  \centering

  \subfloat[Information entropy]{
    \includegraphics[width=0.21\textwidth]{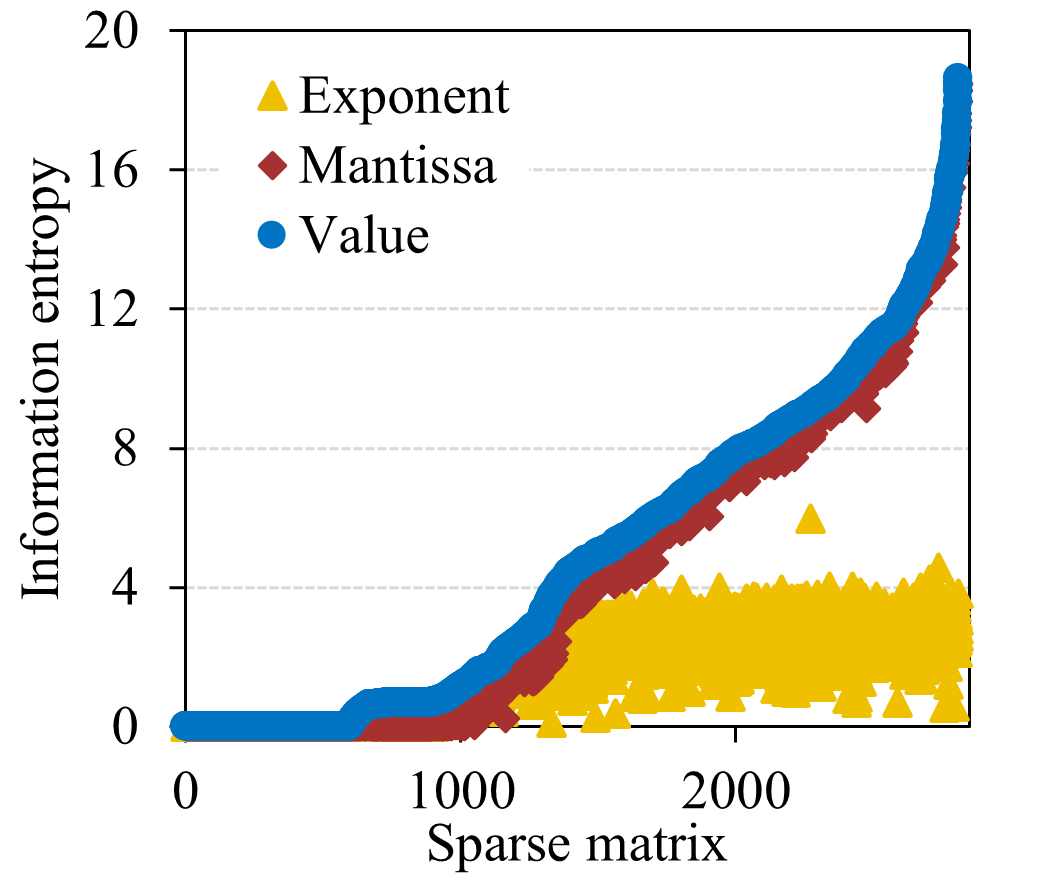}
    }
    \subfloat[Top-1]{
    \includegraphics[width=0.22\textwidth]{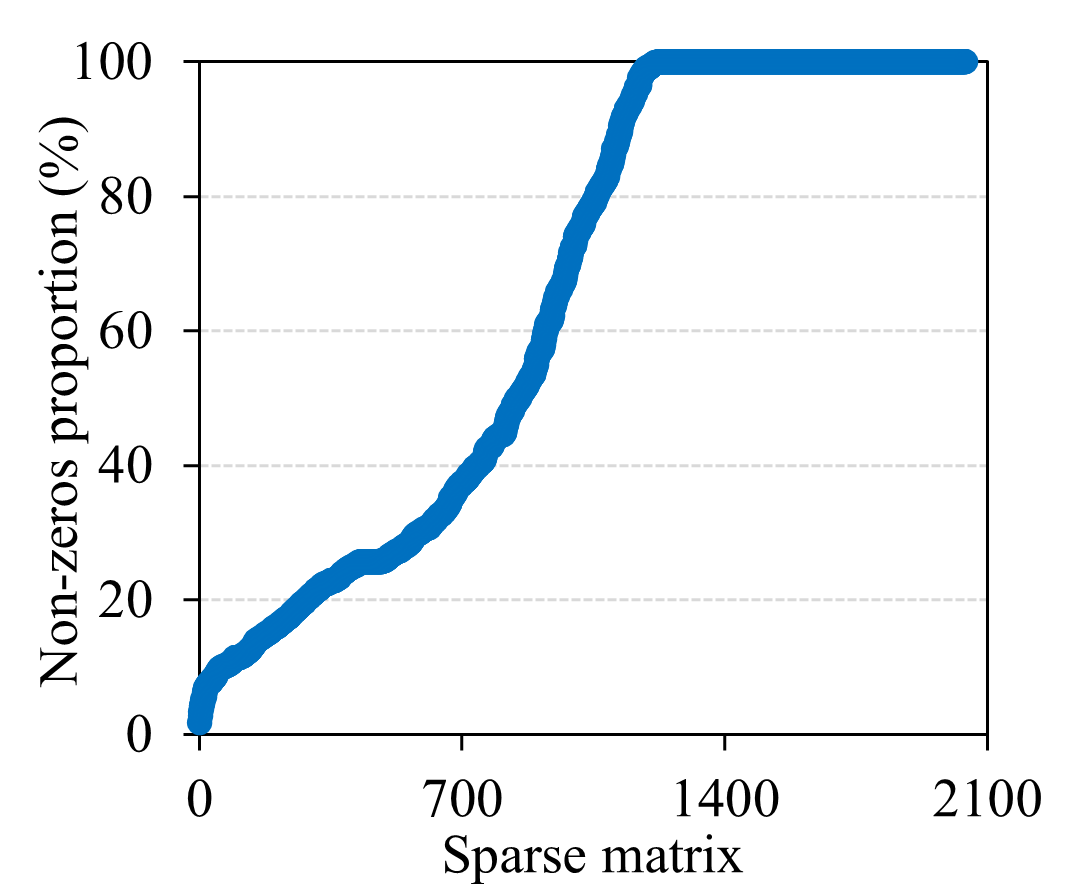}
    }
    \subfloat[Top-2]{
    \includegraphics[width=0.22\textwidth]{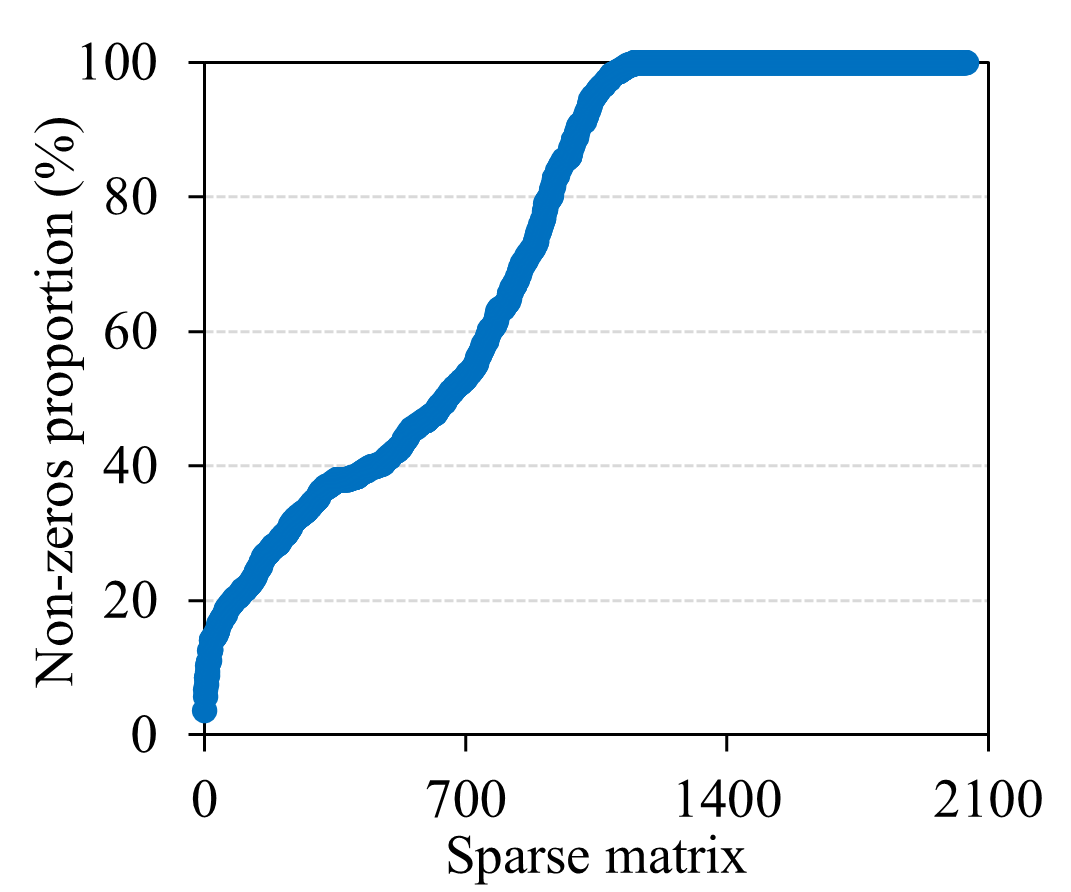}
    }
    \subfloat[Top-4]{
    \includegraphics[width=0.22\textwidth]{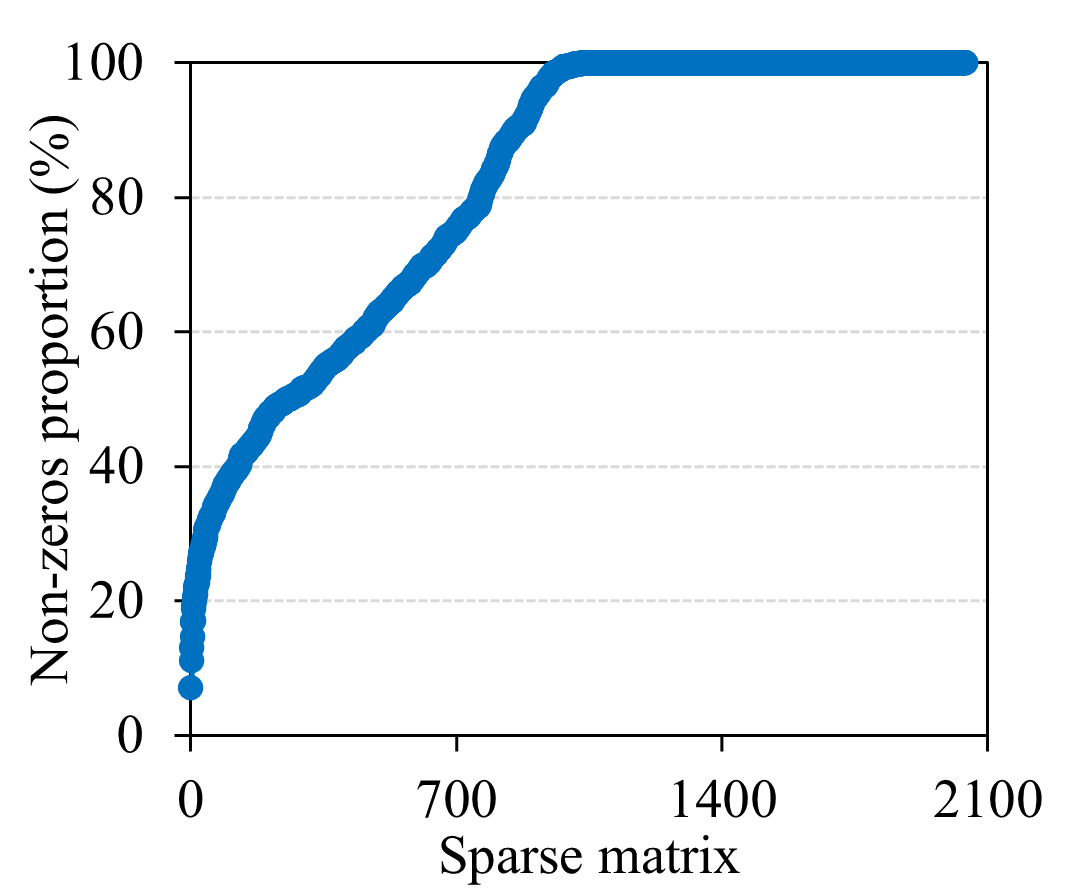}
    }
    \hfill
    \subfloat[Top-8]{
    \includegraphics[width=0.22\textwidth]{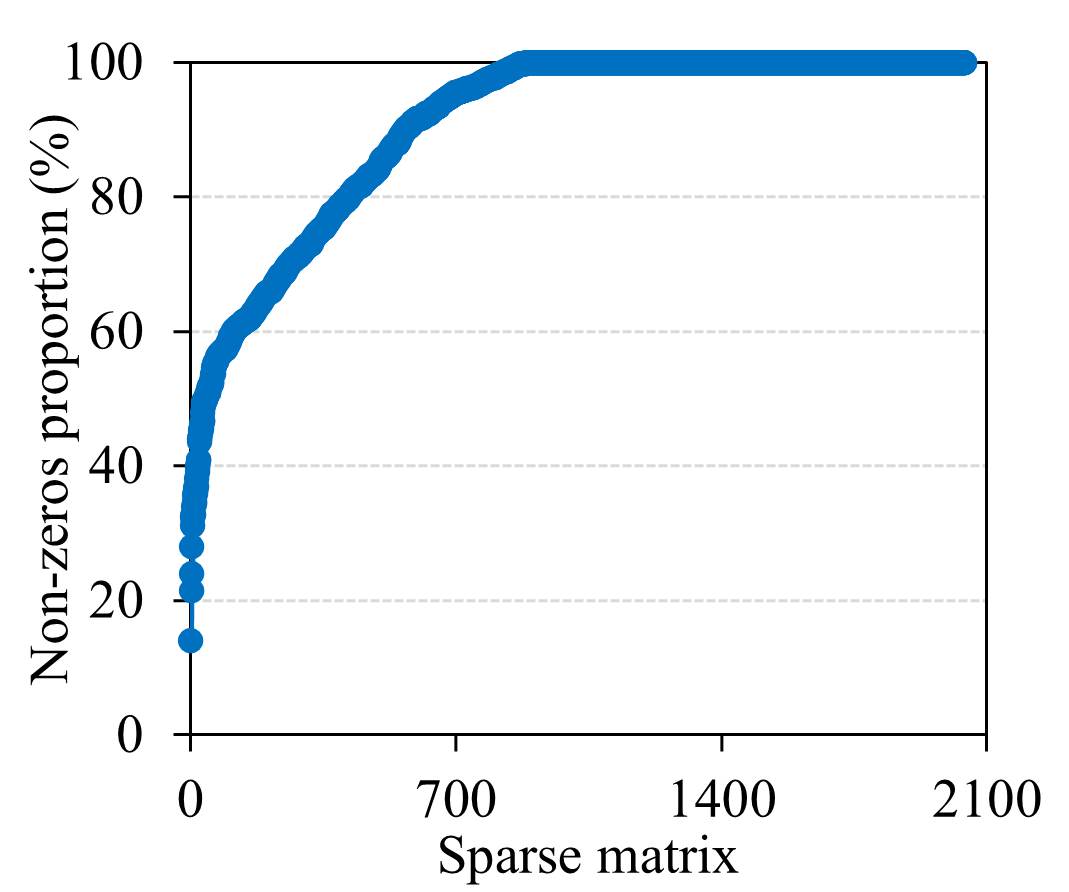}
    }
    \subfloat[Top-16]{
    \includegraphics[width=0.22\textwidth]{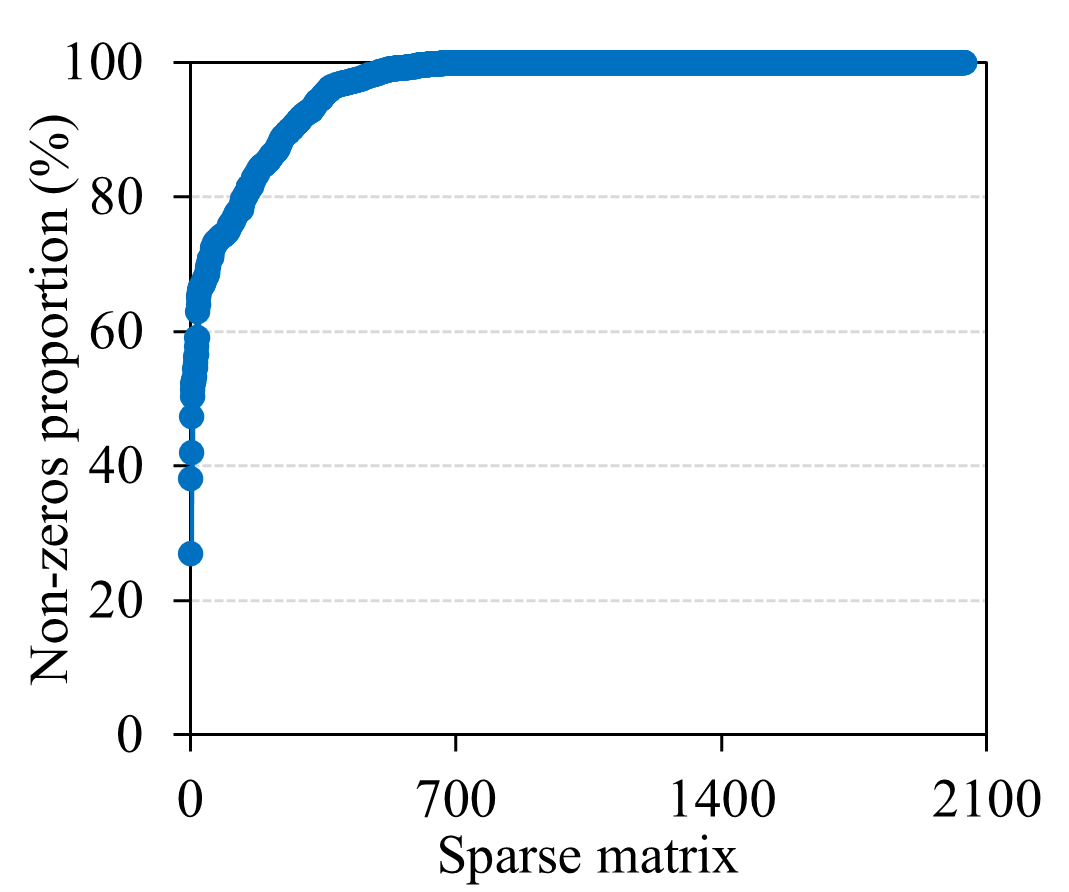}
    }
    \subfloat[Top-32]{
    \includegraphics[width=0.22\textwidth]{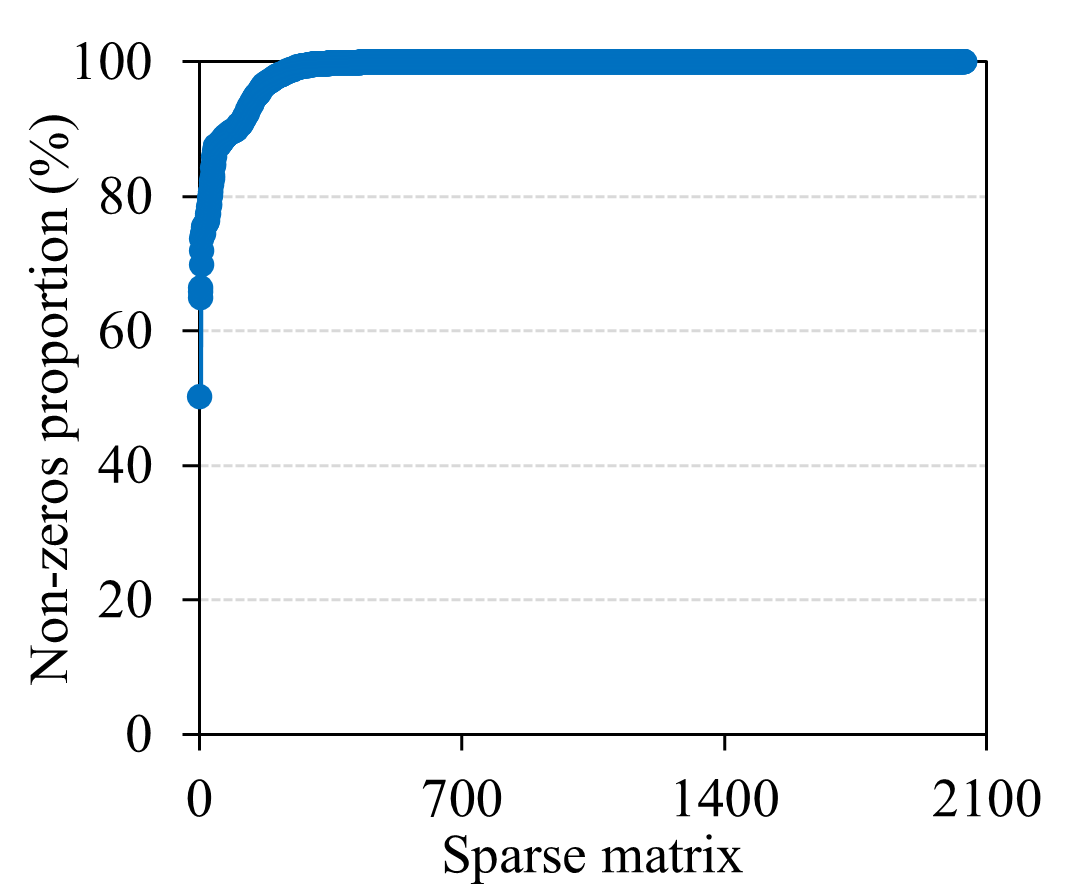}
    }
    \subfloat[Top-64]{
    \includegraphics[width=0.22\textwidth]{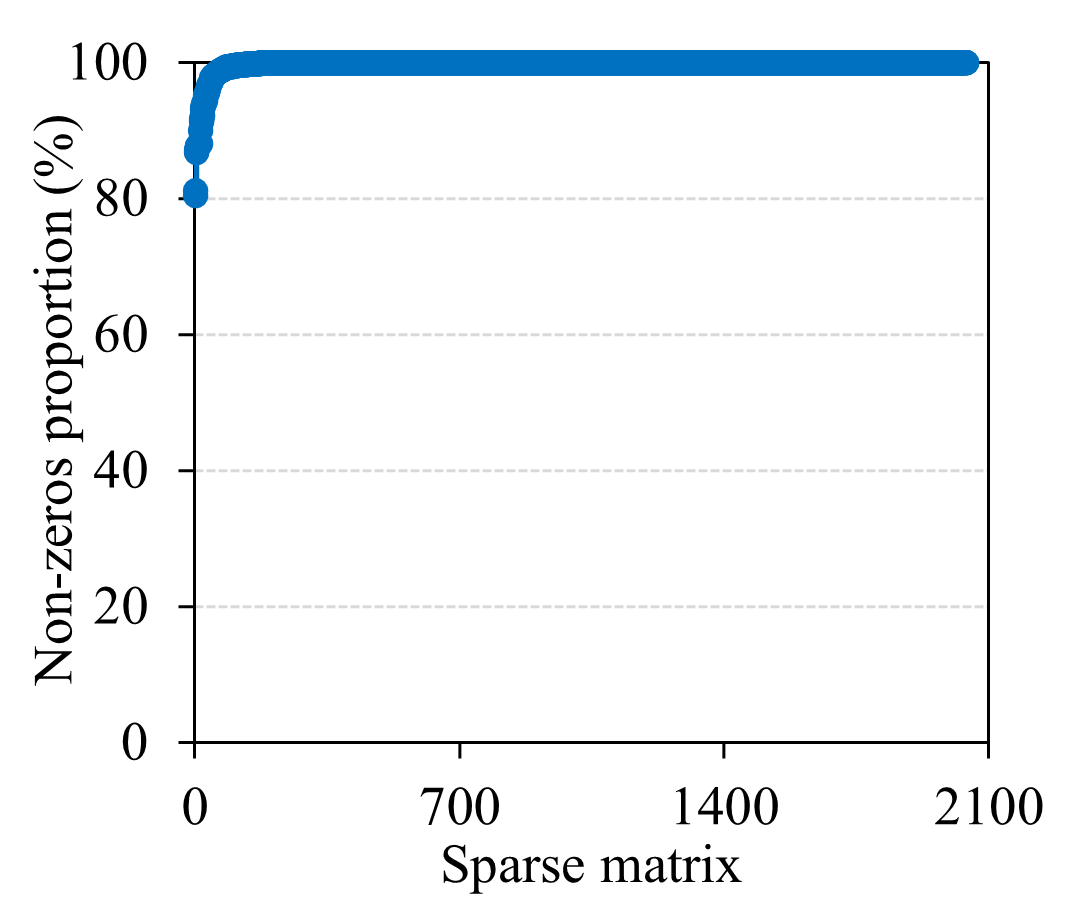}
    }

  \caption{Non-zeros value distribution of sparse matrices. (a) Information entropy of non-zeros values, exponents, and mantissa. (b)-(h) Ratios of non-zeros corresponding to the top-$k$ exponents.}
  \label{fig:non-zero-distribution}
\end{figure*}

Based on the observations above and the observation that most non-zeros in a sparse matrix exhibit close or same exponents, this paper introduces an innovative mixed-precision iterative algorithm based on group-shared exponents. The core idea lies in the extraction and separate storage of shared exponents for a collection of floating-point numbers, thus freeing exponent bits for additional mantissa storage. This method boosts computational efficiency and improves the convergence of iterative computations with lower precision. This approach is practical because the separately stored shared exponents can be represented with more binary bits, preventing data overflow in computations. Moreover, by dedicating all bits, aside from the sign bit, to the mantissa representation, this approach improves the precision of data representation under bit-width constraints. The primary contributions of this paper include:

\begin{itemize}
    \item A new floating-point format is proposed, based on group-shared exponents and segmented storage of the tail, which eliminates the need for storing multiple copies of the same data in different precisions.
    \item A stepped mixed-precision iterative algorithm optimization method based on the group-shared exponents' floating-point format is proposed, which maintains the convergence property of the iterative algorithm while reducing the SpMV memory footprint overhead.
    \item The optimization method for the mixed-precision iterative algorithm introduced in this paper has been validated on a GPU platform. Experimental results demonstrate that our proposed method improves the solving efficiency of the iterative algorithm and achieves similar residuals to high-precision iterative solvers.
\end{itemize}

The remainder of this paper is organized as follows. Section~\ref{sec:motivation} analyzes the numerical distribution of sparse matrices used in iterative algorithms. Section~\ref{sec:method} describes our proposed new floating-point format and stepped mixed-precision iterative algorithm. Section~\ref{sec:evaluation} evaluates the performance of the proposed approach. Section~\ref{sec:related-work} discusses related work, and Section~\ref{sec:conclusion} concludes the paper.

\section{Motivation}\label{sec:motivation}
We have analyzed the numerical characteristics of the non-zeros for over 2,000 sparse matrices from the SuiteSparse Matrix Collection \cite{Davis2011_SuiteSparse}. These matrices come from practical problems in fields such as circuit simulation, linear programming, computational fluid dynamics, and others. We analyzed the values, exponents, and mantissa of non-zeros in these sparse matrices. Figure \ref{fig:non-zero-distribution}(a) presents the distribution of their information entropy, which is calculated using Equation \ref{equ:entropy}, where $p_i$ represents the probability of occurrence. A higher entropy indicates greater uncertainty.

\begin{equation}\label{equ:entropy}
    H(U)=E[-logp_i]=-sum^n_{i=1}(p_i \times logp_i)
\end{equation}

From Figure \ref{fig:non-zero-distribution}(a), it can be observed that the information entropy of non-zeros is very close to that of their mantissas. This is due to the fact that the mantissa of a floating-point number determines its numerical precision. For more than 52\% of the sparse matrices, the information entropy of the non-zero values is greater than 4, while the information entropy of the exponent part is less than 4 for 97\% of the sparse matrices. This indicates that for most of the sparse matrices, the mantissa fields of their non-zeros vary, but the exponent fields exhibit a clustered distribution.

\begin{figure*}[!htbp]
    \centering
    \includegraphics[width=0.75\linewidth]{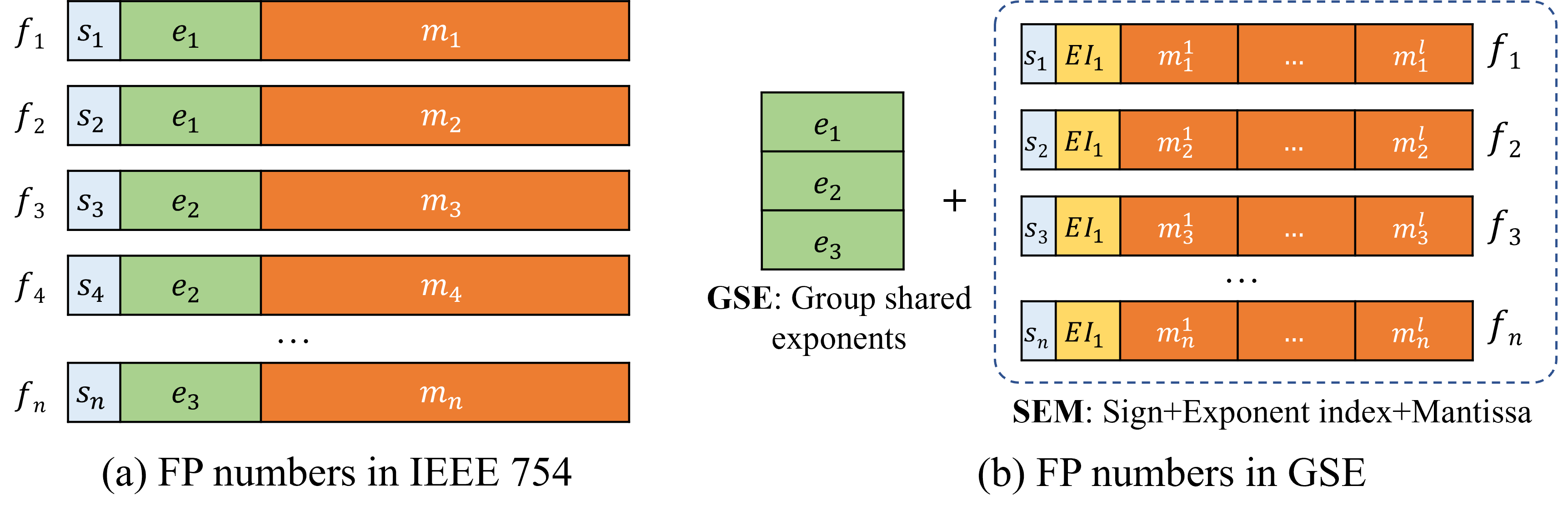}
    \caption{Overview of group-shared exponent format.}
    \label{fig:gce}
\end{figure*}

\begin{equation}
    \label{equ:topk}
    top_k=\sum_{i=1}^{k} NNZ_i/{NNZ}
\end{equation}

We further analyzed the exponent fields of non-zeros by counting the number of non-zeros (NNZ) associated with different exponents in each sparse matrix. Let $NumExp$ denote the number of different exponents in a sparse matrix, and let $S=(e_i,NNZ_i)$ represent the sequence obtained by arranging the non-zeros according to the NNZ with the same exponent, $NNZ_i$, in descending order $(i=1,2,\ldots,NumExp)$. To identify the most frequently occurring exponents, we define ``top-k'' as the $k$ exponents with the highest occurrence, and the non-zeros corresponding to ``top-k'' are calculated using Equation \ref{equ:topk}, where $NNZ$ represents the total NNZ in the sparse matrix. Figures \ref{fig:non-zero-distribution}(b)-(h) show the proportions of non-zeros for ``top-1'', ``top-2'', ``top-4'', ``top-8'', ``top-16'', ``top-32'', and ``top-64'', with their average proportions being 64.7\%, 73.1\%, 82.4\%, 90.9\%, 96.5\%, 98.9\%, and 99.8\%, respectively. These findings suggest that the non-zeros in most sparse matrices share the same or only a few exponents, indicating the under-utilization of the exponent fields' binary bits. By extracting and separately storing the shared exponents, the space originally occupied by the exponent parts can be utilized to store additional mantissa bits, thereby improving precision while reducing memory footprint.

\section{Methodology}\label{sec:method}
\subsection{Overview}
Traditional general-purpose floating-point formats face challenges such as high conversion overhead and tight coupling of storage and computation, which hinder the mixed-precision optimization of iterative algorithms. This paper addresses these challenges by analyzing the numerical characteristics of floating-point numbers within iterative algorithms and proposing an optimization method rooted in the concept of group-shared exponents. Initially, we introduce a novel floating-point representation method based on group-shared exponents. Building on this foundation, we develop a stepped mixed-precision optimization method.

\subsection{Group-Shared Exponent-Based Floating Point Format}
The proposed floating-point format based on group-shared exponents, as depicted in Figure \ref{fig:gce}(b), is a key contribution of this paper. It has two components: the group-shared exponents (GSE) and the sign, exponent index, and mantissa (SEM). The GSE part includes a collection of shared exponents, while the SEM part includes the sign, exponent index, and mantissa of each floating-point number in the set $F$. Notably, the number of shared exponents, represented as $k$, is typically much smaller than the total number of floating-point numbers ($n$). Storing exponents separately eliminates the redundant storage of exponents. This approach reduces memory requirements and improves the efficiency of exponent storage.

\subsubsection{Shared Exponents Extraction}
For a given set $F$ of floating-point numbers and the number of shared exponents $numExp$, firstly, it is necessary to find the shared exponents. For a set of floating-point numbers, we first count the occurrences $N_i$ for each distinct exponent $e_i$. Subsequently, the exponents are sorted in descending order based on their occurrence counts to identify the $k$ most frequent exponents. These exponents collectively constitute the GSE part. For a given application targeting a specific area, the distribution of floating-point numbers has a certain degree of stability. After finding the shared exponents that maximize the coverage through a single-pass analysis, the group exponent setting can be reused in subsequent calculations without unnecessary reanalysis to obtain a new group of shared exponents. Additionally, for applications demanding low preprocessing overhead, the shared exponents can be calculated using sampling techniques. For instance, a sparse matrix is divided into several row blocks, and the exponents' distribution in a random row is calculated for each row block, serving as the exponents' distribution in that block.

Separating the exponent and mantissa parts of floating-point numbers reduces redundancy in exponent storage and enables an increased bit-width for the mantissa representation. However, it is necessary to obtain the exponent part of each floating-point number during computations, concatenate it with the mantissa part, and perform calculations. Modern processors only support arithmetic operations based on the general-purpose floating-point formats. To address this challenge, a small portion of the bit-width in the SEM part is used to label the corresponding exponent, known as the exponent index. The number of shared exponents determines the bit-width of the exponent index part. Therefore, the number of shared exponents should be manageable in size; otherwise, the number of binary bits representing the mantissa part decreases.


\subsubsection{Denormalized Representation}

The limited shared exponents may not represent all data in a floating-point set. To extend the representation range, we use the denormalized representation of the mantissa in our proposed format. Existing floating-point numbers use the normalized representation. There is an implicit, hidden bit in the mantissa of these floating-point numbers. This bit is present virtually in the mantissa but not stored in memory because it is always 1. Therefore, in our proposed format, we increment all the shared exponents by~1 to explicitly represent the hidden~1. To convert all non-zeros in the floating-point set to the group-shared exponent-based format, one of the shared exponents must be the maximum exponent of all non-zeros plus one; otherwise, a few non-zeros may not be represented.

Here, we consider converting a double-precision floating-point vector $vecD$ to GSE-SEM format $vecSEM$, where the SEM part occupies a bit-width of 16, for example. Algorithm~\ref{alg:formatConvert} presents the format conversion process, where $EI\_bit$ represents the bit-width for storing exponent indexes, $N$ is the vector size, $numExp$ represents the number of shared exponents, and the $SEM$ array stores those shared exponents. First, we extract the sign bit and exponent field from the 64-bit floating-point number $valD$ (lines 2-4). Then, we check if the exponent exists in the array $SEM$ (lines 6-12). If it does not exist, we find the nearest shared exponent greater than the exponent $exp$ and calculate the difference $minDiff$ between them for subsequent denormalization (lines 13-21). Next, we shift the exponent index $expIdx$ to the binary bit where exponent indexes are stored in the 16-bit SEM (line~22). Then, we denormalize the mantissa using \textit{minDiff} (lines 23-25), where $MAX\_52$ represents a number in which the least significant 52 bits are all 1 and the rest are all 0. Finally, we combine the sign, exponent index, and mantissa using logical OR operations and get the converted SEM storage $vecSEM[tid]$ (line~26).

\begin{algorithm}
        \KwIn{$vecD$, $N$, $numExp$, $SEM$, $EI\_bit$}
        \KwOut{$vecSEM$}
        \For{($i=0; i<N; i++$)}{
            $valD = vecD[i]$; \\
            $sign = (valD >> 48) \& 0x8000$; \\
            $exp = (valD >> 52) \& 0x7FF$; \\
            $minDiff = UINT\_MAX$; \\
            \For{($k = 0; k < numExp; k++$)}{
                \If{($exp + 1 == SEM[k]$)}{
                    $expIdx = k$; \\
                    $minDiff = 1$;\\
                    $break$;\\
                }
            }
            \If{($k == numExp$)}{
                \For{($jj = 0; jj < numExp; jj++$)}  {
                    $diff = SEM[jj] - exp$; \\
                    \If{($diff > 0\textup{ }\&\&\textup{ }diff < minDiff$)}{
                        $minDiff = diff$; \\
                        $expIdx = jj$; \\
                    }
                }
            }
            $expIdx = expIdx << (15-EI\_bit)$;\\
            $valD = (valD \& MAX\_52) >> minDiff$; \\
            $valD = valD >> (37+EI\_bit)$; \\
            $valD = valD | (0x1 << (15 - EI\_bit - minDiff))$; \\
            $vecSEM[i] = sign | expIdx | valD $; \\
        }
        \caption{Pseudocode for converting double-precision vector to GSE-SEM vector.}
        \label{alg:formatConvert}
\end{algorithm}

\subsubsection{Segmented Mantissa Storage}
\begin{figure}[!htbp]
    \centering
    \includegraphics[width=0.99\linewidth]{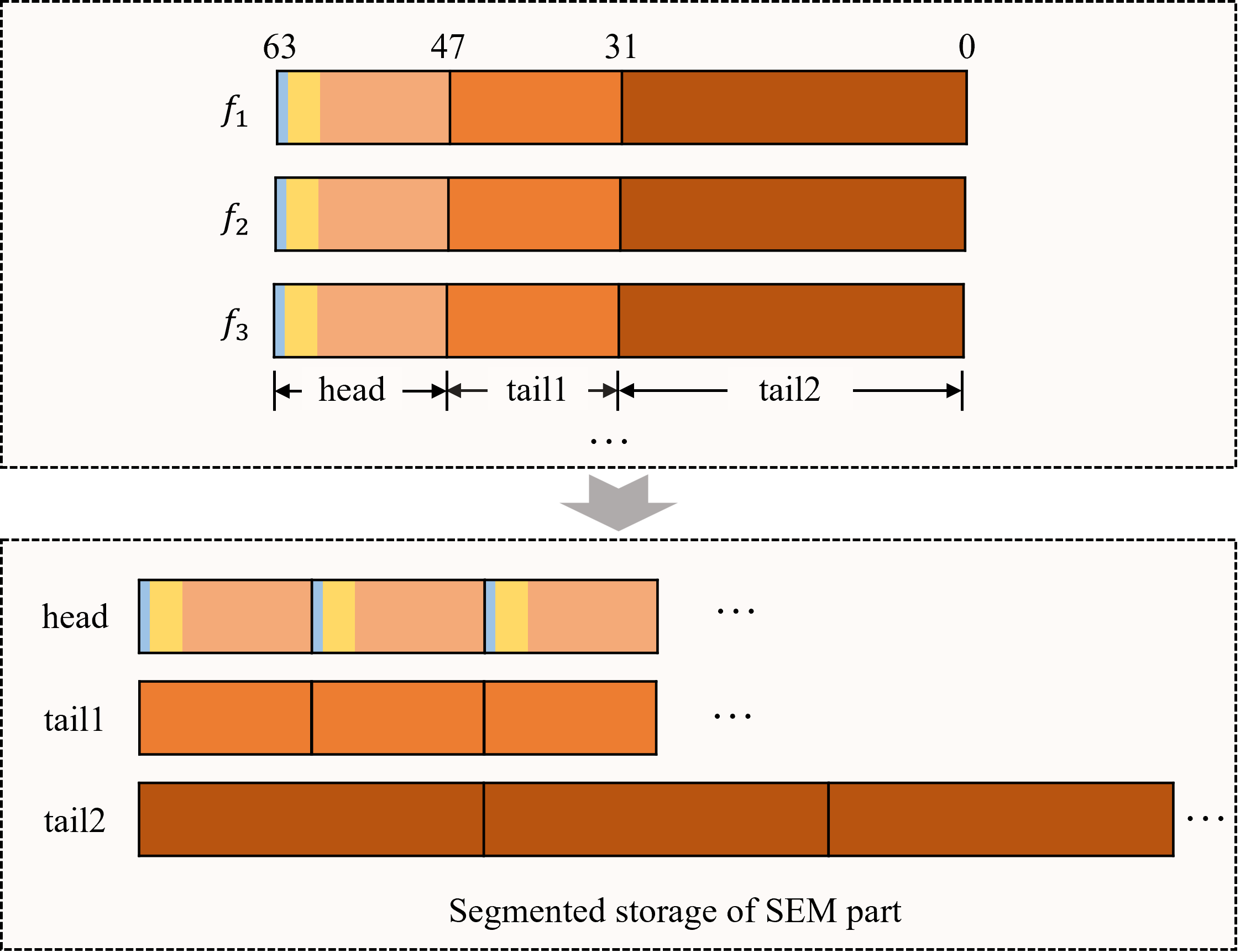}
    \caption{Segmented storage of SEM.}
    \label{fig:segmented-SEM}
\end{figure}

Due to the varying precision requirements of different applications, application optimization based on mixed-precision techniques requires the conversion between different floating-point formats. Moreover, maintaining different precisions of the same data in memory results in storage waste. This paper uses segmented storage for the SEM part of the GSE-SEM format. In \cite{DBLP:journals/concurrency/GrutzmacherCFGA20}, a 64-bit floating-point number is divided into two segments. The first segment consists of the most significant 32 bits, and the remaining least significant 32 bits form the second segment. In our SEM part, taking the 64-bit SEM as an example, as shown in Figure \ref{fig:segmented-SEM}, it is divided into three segments. The first segment contains the most significant 16 bits, including one sign bit, some bits representing the index of the shared exponent, and some mantissa bits, which we call the head shared by all segments. The remaining mantissa bits are divided into two segments: the top 16 bits form the second segment called $tail1$, and the remaining least significant bits form the third segment called $tail2$. As shown in the lower part of Figure \ref{fig:segmented-SEM}, the heads of all floating-point numbers are stored consecutively in memory, followed by $tail1$ and $tail2$ stored in sequence, respectively. This memory layout enables the full utilization of the memory bandwidth and coalesced memory access in computation. If high-precision floating-point representation is required, concatenating the head and tails forms the high-precision representation. That is, the head can be concatenated with $tail1$ or with $tail1$ and $tail2$.

\subsection{GSE-SEM Based SpMV}
\subsubsection{Sparse Matrix Representation}
Sparse matrices have many zero elements and thus are usually stored in compressed formats. The Compressed Sparse Row (CSR) format is one of the most popular compressed formats. It includes three arrays: \textit{RowPtr}, \textit{ColIdx}, and \textit{Val}, which store the index of the first non-zero element in each row, the column index, and the value of each non-zero element, respectively. 32-bit unsigned integers are usually used to store column indices, while the most significant bits are unused in most situations. In the SuiteSparse Matrix Collection \cite{Davis2011_SuiteSparse}, the largest column size is $226,196,185$, and the most significant four bits are unused. Therefore, for sparse matrix representation, the indices of shared exponents can be encoded into the column indices of non-zeros to free up more binary bits in non-zero values for tail representation. When the column size of a sparse matrix is so large that there are not enough binary bits to store the shared exponent indices, we can encode them into the value array.


\subsubsection{SpMV}
Separated storage of shared exponents and segmented storage of tail fields promote the multi-precision representation and switching of sparse matrices, thereby supporting different-precision SpMV calculations based on the same sparse matrix. Considering that low-precision computing impacts the accuracy of SpMV applications, we load low-precision sparse matrices only during memory access and still perform multiplication and accumulation operations based on double-precision floating-point representation. To support mixed-precision iterative calculations, we designed three SpMV operators using different precisions of sparse matrices.

\begin{algorithm}
         \KwIn{$I$, $J$, $V\_h$, $expArr$, $\vect{x}$}
        \KwOut{$\vect{y}$}
        \For{$(i=0; i<rows; i++)$}{
            \For{$j=I[i]; j < I[i+1]; j++$}{
                $col = J[j]$;\\
                $expIdx = col >> 29$;\\
                $col\textup{ }\&= MAX\_29$;\\
                $val = V\_h[j]$; \\
                $val\_FP64 = (val \& 0x8000) << 48$; \\
                $base = 14$; \\
                $pos = \_\_fns(val, base, -1)$; \\
                \uIf{$pos \ne MAX\_32$} {
                    $val\_FP64\textup{ }|=\textup{ }(expArr[expIdx]-(15-pos) << 52)$; \\
                    $temp = (val \& 0x7FFF) << 36$;\\
                    $temp = (temp << (16-pos)) \& MAX\_52$; \\
                    $val\_FP64\textup{ }|=\textup{ }temp$;\\
                }\Else{
                    $val\_FP64 = 0$; \\
                }
                $sum += val\_FP64 \times \vect{x}[col]$; \\
            }
            $\vect{y}[i] = sum$;\\
        }
        \caption{SpMV calculation using only \textit{head} part.}
        \label{alg:spmv}
\end{algorithm}

Algorithm \ref{alg:spmv} presents the SpMV using only the $head$ part, where $MAX\_M$ represents a number in which the least significant $M$ bits are all 1 and the rest are all 0. For each non-zero element of the sparse matrix, we first extract the exponent index $expIdx$ from the column index and rebuild the actual column index $col$ (lines 3-5). Then, we take the sign bit (i.e., the most significant bit) from the 16-bit head $val$ corresponding to the non-zero element and place it in the most significant bit of the 64-bit variable $val\_FP64$ (lines 6-7). Next, we scan the remaining 15 bits of $val$ from the most significant to the least significant bit and find the position $pos$ where the first 1 appears (lines 8-17). If it is not found in the head, i.e., $pos$ equals $MAX\_32$, we consider the non-zero element value small and set it to 0 (line 16). Otherwise, we use the shared exponent $expArr[expIdx]$ and $pos$ to convert the exponent from denormalized to normalized representation and place it in $val\_FP64$ (line 11). Then, we extract the mantissa part from $val$, convert it from denormalized to normalized representation (lines 12-13), and place it in $val\_FP64$ (line 14).

In SpMV calculation, we use both $head$ and $tail1$; in addition to loading $head$, we also need to load $tail1$. The final concatenated $val\_FP64$ includes more mantissa bits than $val\_FP64$ in Algorithm~\ref{alg:spmv}, providing higher precision. Similarly, the non-zero elements with the highest precision are loaded in SpMV calculation using $head$, $tail1$, and $tail2$. Therefore, three-precision SpMV operators based on the GSE-SEM format are provided.

\subsection{Stepped Mixed-Precision Iterative Solvers}
Multiple precision SpMV operators provide the possibility for mixed-precision optimization of iterative algorithms. In actual hardware, the finite precision of floating-point numbers leads to the non-associativity of floating-point arithmetic. Hence, it is challenging to theoretically ensure whether iterative algorithms could converge when using different precisions. Considering that the convergence property of iterative algorithms is highly related to the coefficient matrix of linear equations, this paper proposes a stepped mixed-precision optimization method based on the proposed GSE-SEM. The method starts iterations with a lower-precision sparse matrix read from memory and monitors the change of residuals during the iteration. When the change in residuals slows down, more mantissa bits are read from memory to concatenate with the head into a higher-precision sparse matrix for continuing iteration, and so forth. The algorithm continues with the higher-precision sparse matrix to calculate a satisfactory approximate solution. This mixed-precision optimization algorithm fully leverages the advantages of the GSE-SEM format, exploring the optimization space by reading different precisions of sparse matrices within iterative algorithms. The GSE-SEM-based mixed-precision iterative algorithm improves the binary bit utilization in floating-point representation by extracting redundant common exponents and storing them separately. Additionally, this algorithm enables GMRES calculations of varying precisions by occupying only the storage space of a high-precision sparse matrix, thereby decoupling storage from the computation.

The workflow of the stepped mixed-precision iterative algorithm is presented in Algorithm \ref{alg:stepped-GMRES}. Before the iteration begins, we pre-store the sparse matrix in the GSE-SEM representation to support the different-precision SpMV calculations during the iteration. To simplify the algorithm, we use $\matr{A}_1$, $\matr{A}_2$, and $\matr{A}_3$ to represent three sparse matrices with different precisions. It is worth noting that these sparse matrices are generated by concatenating the head and tails during the iterative SpMV computation without redundant storing. During the iteration process, we monitor the residual changes and choose whether an increase in precision is necessary. If necessary, we modify the precision label ($tag$) to call a higher-precision SpMV operator.


\begin{algorithm}
         \KwIn{$\matr{A}_1$ ($head$ only), $\matr{A}_2$ ($head$ + $tail1$), $\matr{A}_3$ ($head$ + $tail1$ + $tail2$)}
        \KwOut{approximate solution $\hat{\vect{x}}$}
        $j=1$; $tag = 1$; \\
        \While{$<$Stopping criteria not satisfied$>$}
        {
            \uIf{$tag == 1$}
            {
                $\vect{w}_j=\matr{A}_1\vect{v}_j$;
            }
            \uElseIf{$tag == 2$}{
                $\vect{w}_j=\matr{A}_2\vect{v}_j$;
            }
            \Else{
                $\vect{w}_j=\matr{A}_3\vect{v}_j$;
            }
             ......\tcp*[h]{Other operations are omitted} \\
            \If{$<$higher precision is required$>$}
            {
                \uIf{$tag == 1$}{
                    $tag = 2$;
                }
                \uIf{$tag == 2$}{
                    $tag = 3$;
                }
            }
            $j++$;
        }
        \caption{Stepped mixed-precision iterative algorithm.}
        \label{alg:stepped-GMRES}
\end{algorithm}



We choose whether to switch precision by monitoring the residuals in recent iterations. Specifically, in the first $l$ iterations, the low-precision sparse matrix is read for computation, while the residuals from the past $t$ iterations are recorded, and $t<l$. Afterward, for every $m$ iteration, the following three metrics are calculated. Let $j$ represent the current iteration:

1) \textit{RSD}: The relative standard deviation of the residuals from the past $t$ iterations, as shown in Equation (\ref{equ:evp}), where $resid[i]$ represents the residual of the $i$-th iteration

\begin{equation}\label{equ:evp}
    RSD = \frac{\sqrt{\frac{1}{t} \sum_{i=j-t}^{j-1} (resid[i] - avg)^2}}{avg}
\end{equation}

2) \textit{nDec}: The number of times the residual has decreased over the past $t$ iterations, as shown in Equations (\ref{equ:desc}) and (\ref{equ:desc-comp}).
\begin{equation}\label{equ:desc}
    nDec = \sum_{i=j-t}^{j-1} f(i)
\end{equation}
\begin{equation}\label{equ:desc-comp}
    f(i) = 
    \begin{cases}
    1& resid[i] \textgreater resid[i+1]\\
    0& resid[i] \leq resid[i+1]
    \end{cases}
\end{equation}

3) \textit{relDec}: The relative decrease rate of residuals over the past $t$ iterations, as shown in Equation (\ref{equ:comp}).
\begin{equation}\label{equ:comp}
    relDec = \frac{resid[j-t]-resid[j-1]}{resid[j-t]}
\end{equation}




Three conditions for deciding whether to increase the precision are proposed based on these metrics.

\textbf{Condition 1}: $RSD > RSD\_limit$ \&\& $nDec < t/2$. $RSD\_limit$ is a predefined threshold. Satisfying this condition means that the residuals are difficult to decrease, and there are significant fluctuations in the residuals of the past $t$ iterations.

\textbf{Condition 2}: $nDec \geq t/2$ \&\& $relDec < relDec\_limit$. $relDec\_limit$ is a predefined threshold. Satisfying this condition means that the rate of residual decrease in the past $t$ iterations was slower than that in the previous $t$ iterations.

\textbf{Condition 3}: $nDec = 0$. Satisfying this condition means the residual has not decreased during the past $t$ iterations.

If any of these three conditions are met, the higher-precision SpMV operator is called in the subsequent iterations.

\section{Evaluation}\label{sec:evaluation}
\subsection{Experimental Setup}
Our experiments were carried out on a machine with a NVIDIA V100-SXM2 32 GB GPU. Table \ref{tab:environment-setup} presents the hardware and software configuration of the machine. In the subsequent experimental evaluation, all SpMV implementations are based on the CSR-Vector algorithm of the CUSP library \cite{Cusp}, where the number of threads allocated per row is set according to the decision tree proposed in \cite{DBLP:conf/icpads/GaoJLSWS21} for each algorithm. The runtime of SpMV is the average time taken over 100 executions. In addition, in SpMV calculations, the elements of the multiplication vector are set to 1 to observe the errors caused by low-precision sparse matrix representation. All vector operations in the iterative algorithms are performed by calling APIs in the NVIDIA cuBLAS library. For both iterative algorithms, CG and GMRES, the residual threshold is set to $10^{-6}$. In GMRES, the restart is set to 30, and the maximum outer iterations are set to 500. In other words, the maximum iterations for GMRES are 15,000. For the CG solver, the maximum iterations are set to 5,000.

\begin{table}[!htbp]
    \centering
    \caption{Hardware and software.}
    \scalebox{0.9}{
    \begin{tabular}{lcl}
    \toprule
        \multirow{4}{*}{\centering \textbf{Hardware}} & 
        \multirow{2}{*}{\centering Host} & 
            \textbf{CPU}: Intel Xeon Platinum 8255C, 2.50 GHz, 40 cores \\
        & & \textbf{Memory}: 128 GB  \\ 
       \cmidrule{2-3}
        & \multirow{2}{*}{Device} &
            \textbf{GPU}: V100-SXM2, 5120 cores \\
        & & \textbf{Memory}: 32 GB, B/W 898 GB/s \\ 
        \midrule
        \multirow{4}{*}{\centering \textbf{Software}} & 
        OS & 64-bit CentOS 7.9 \\ \cmidrule{2-3}
        & Compiler &  
        \textbf{nvcc}: 12.0, \textbf{gcc/g++}: 8.3.0 \\ \cmidrule{2-3}
        & Library & 
       CUDA: 12.0, 
       CUSP:0.5.1\\ 
       \bottomrule
    \end{tabular}} 
     \label{tab:environment-setup}
\end{table}

All tested sparse matrices are from the well-known SuiteSparse Matrix Collection \cite{Davis2011_SuiteSparse}. The test set used to evaluate the number of shared exponents $k$ and SpMV performance includes 312 sparse matrices. Their number of non-zeros ranges from 10 to $10^8$. For the experimental evaluation of typical iterative algorithms, we use different test sets. Considering that GMRES is designed for solving asymmetric sparse linear systems, we selected 15 asymmetric sparse matrices to form the test set for GMRES. The CG algorithm is designed for solving symmetric positive definite sparse linear systems, so we selected an additional 15 sparse matrices to form the CG test set. Table \ref{tab:matrix-list} presents the details of the two test sets.

\begin{table*}[!ht]
    \centering
    \caption{Test sets of CG and GMRES.}
    \begin{tabular}{clrr|clrr}
    \toprule
    \multicolumn{4}{c|}{Test set of CG} & \multicolumn{4}{c}{Test set of GMRES} \\ \midrule
    ID & Matrix & Rows & Number of non-zeros & ID & Matrix & Rows & Number of non-zeros \\ \midrule
        1 & bcsstk09 & 1,083  & 18,437  & 1 & iprob &  3,001  &  9,000  \\
        2 & bcsstm24 & 3,562  & 3,562  & 2 & dw1024 &  2,048  &  10,114  \\ 
        3 & bundle1 & 10,581  & 770,811  & 3 & dw2048 &  2,048  &  10,114  \\ 
        4 & ted\_B & 10,605  & 144,579  & 4 & adder\_dcop\_01 &  1,813  &  11,156  \\
        5 & cvxbqp1 & 50,000  & 349,968  & 5 & init\_adder1 &  1,813  &  11,156  \\ 
        6 & consph & 83,334  & 6,010,480  & 6 & adder\_dcop\_39 &  1,813  &  11,246  \\ 
        7 & m\_t1 & 97,578  & 9,753,570  & 7 & Pd &  8,081  &  13,036  \\ 
        8 & Dubcova3 & 146,689  & 3,636,643  & 8 & add32 &  4,960  &  19,848  \\
        9 & af\_0\_k101 & 503,625  & 17,550,675  & 9 & TS &  2,142  &  45,262  \\
        10 & af\_1\_k101 & 503,625  & 17,550,675  & 10 & epb2 &  25,228  &  175,027  \\ 
        11 & af\_shell4 & 504,855  & 17,562,051  & 11 & wang3 &  26,064  &  177,168  \\ 
        12 & Fault\_639 & 638,802  & 27,245,944  & 12 & 3D\_28984\_Tetra &  28,984  &  285,092  \\ 
        13 & bone010 & 986,703  & 47,851,783  & 13 & raefsky1 &  3,242  &  293,409  \\ 
        14 & thermal2 & 1,228,045  & 8,580,313  & 14 & atmosmodl &  1,489,752  &  10,319,760  \\ 
        15 & Queen\_4147 & 4,147,110  & 316,548,962  & 15 & ML\_Geer &  1,504,002  &  110,686,677 \\ \bottomrule
    \end{tabular}
    \label{tab:matrix-list}
\end{table*}


\begin{figure*}
  \centering
  \subfloat[Time speedup]{
    \includegraphics[width=0.48\textwidth]{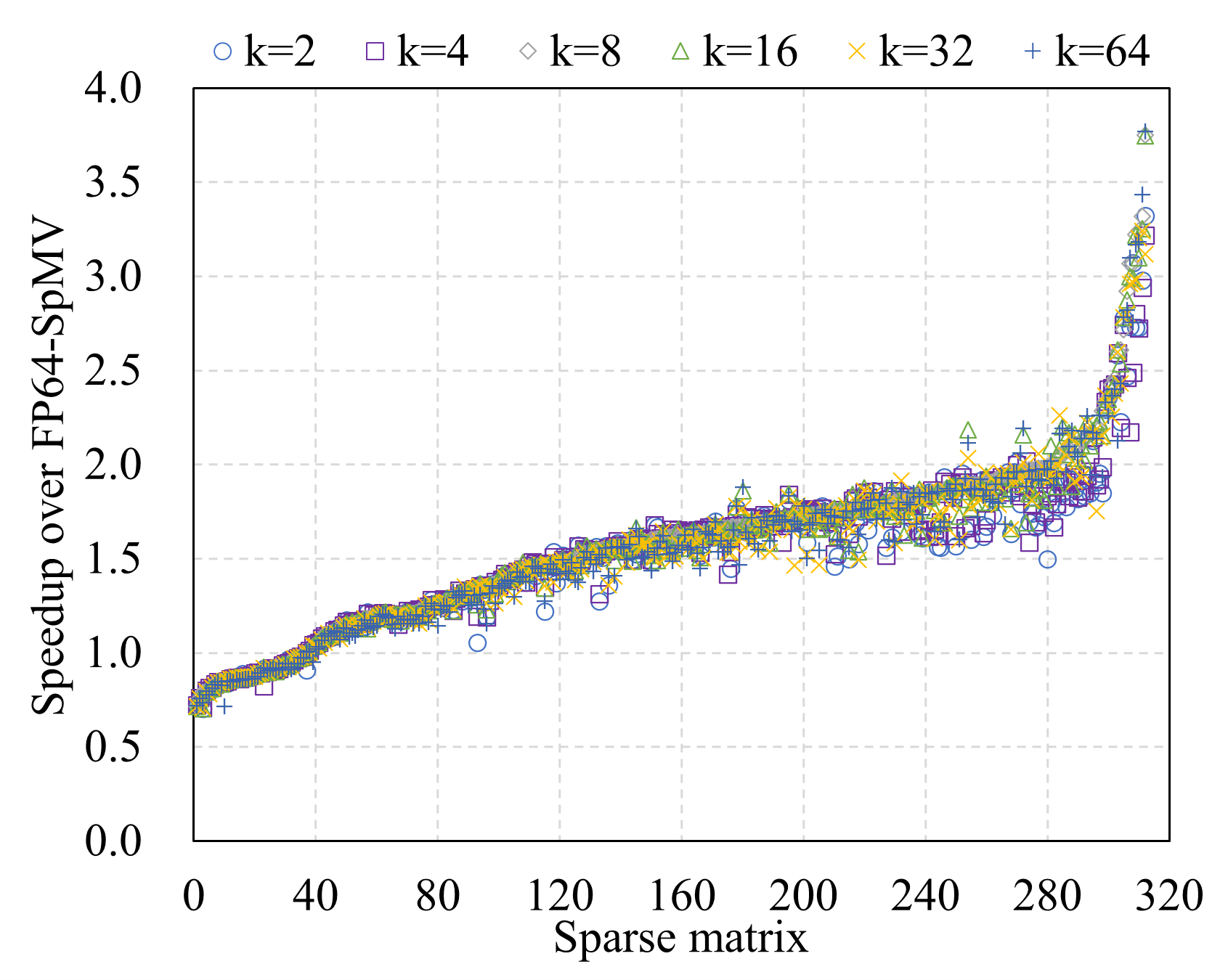}
    }
    \subfloat[Maximum absolute error]{
    \includegraphics[width=0.48\textwidth]{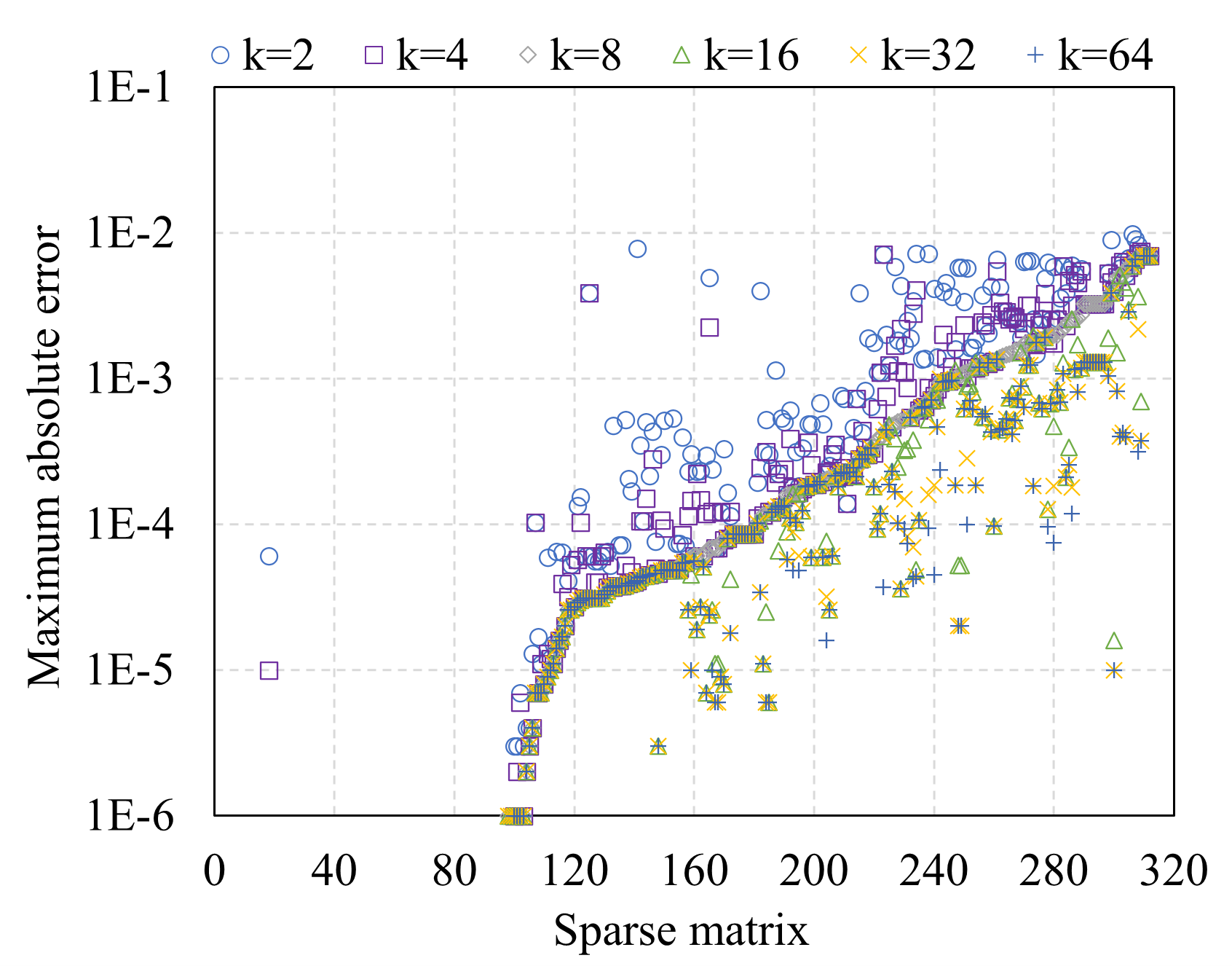}
    }
  \caption{Performance and error comparison under different numbers of shared exponents $k$ for more than 300 sparse matrices. (a) Speedup of the GSE-SEM-SpMV to the FP64-SpMV under different $k$. (b) The maximum absolute error (labeled as maxAbsErr) between the result vector of GSE-SEM-based SpMV and that of the FP64-SpMV under different $k$.}
  \label{fig:shared_exponents}
\end{figure*}

\subsection{Number of Shared Exponents}\label{sec:N-shared-exponents}
\begin{figure}[!htbp]
    \centering
    \includegraphics[width=0.99\linewidth]{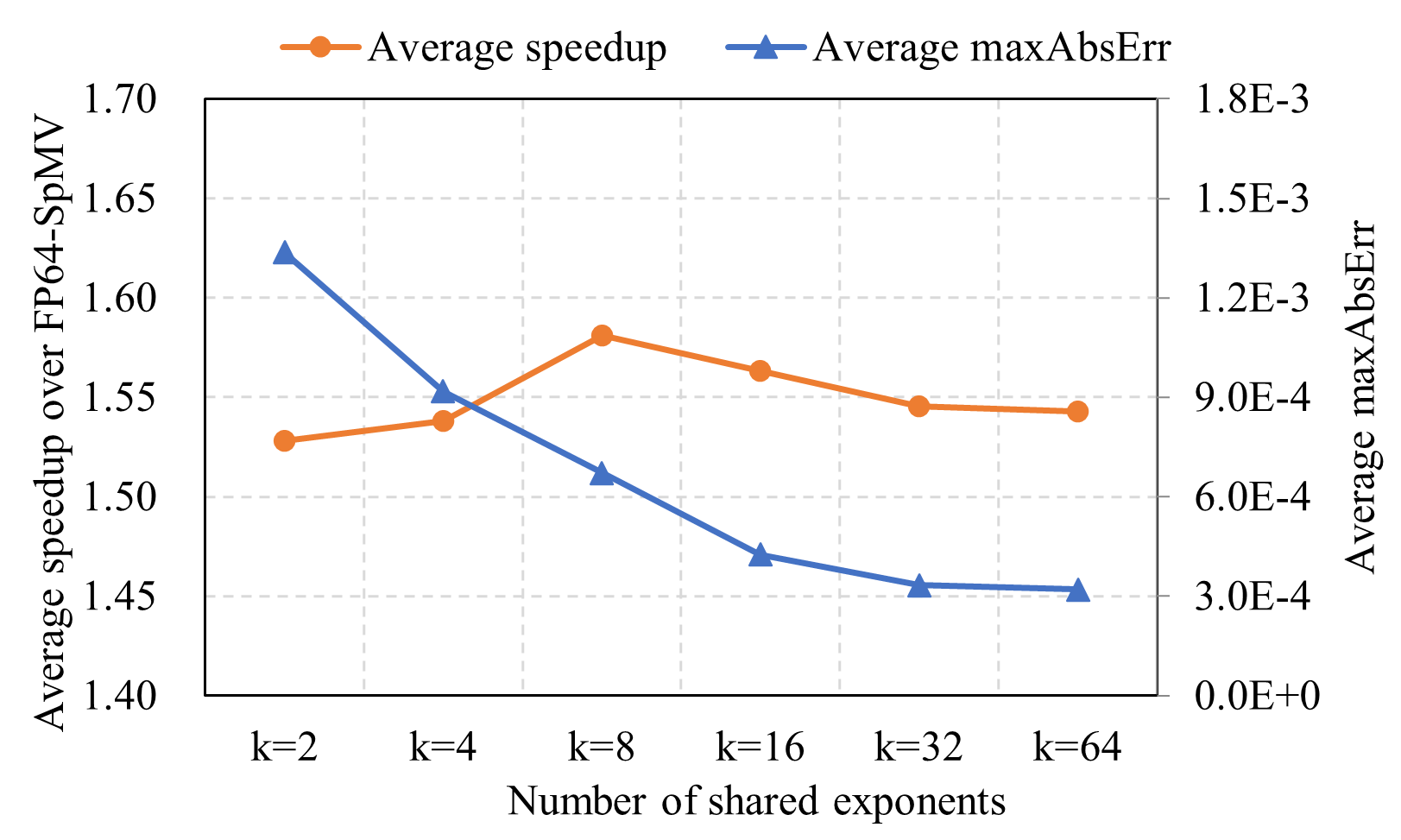}
    \caption{The average speedups of the GSE-SEM-SpMV to the FP64-SpMV under different numbers of shared exponents $k$.}
    \label{fig:mean-spmv-speedup}
\end{figure}

The optimal number of shared exponents varies for different sparse matrices. If there are too many shared exponents, and in extreme cases, each floating-point number has its special exponent, the GSE-SEM floating-point representation becomes equivalent to the traditional IEEE 754 floating-point representation, and this method cannot bring performance benefits. Conversely, if the number of shared exponents is too small, the precision loss caused by denormalized representation increases. We evaluated the impact of the number of shared exponents ($k$) on the performance of the GSE-SEM-based SpMV using more than 300 sparse matrices. $k$ is set to 2, 4, 8, 16, 32, and 64, and the GSE-SEM-based SpMV only uses the head part. Figure \ref{fig:shared_exponents} compares the time and the calculated results. Figure \ref{fig:shared_exponents}(a) presents the time speedup of GSE-SEM-SpMV using the head part relative to the FP64-based SpMV (FP64-SpMV) at different $k$, and Figure \ref{fig:shared_exponents}(b) presents the maximum absolute error between the calculated vectors of FP64-SpMV and GSE-SEM-SpMV. The results are sorted in ascending order according to the result of $k$=8. Their averages are presented in Figure~\ref{fig:mean-spmv-speedup}.

\begin{figure*}
  \centering
  \subfloat[GFLOPS]{
    \includegraphics[width=0.48\textwidth]{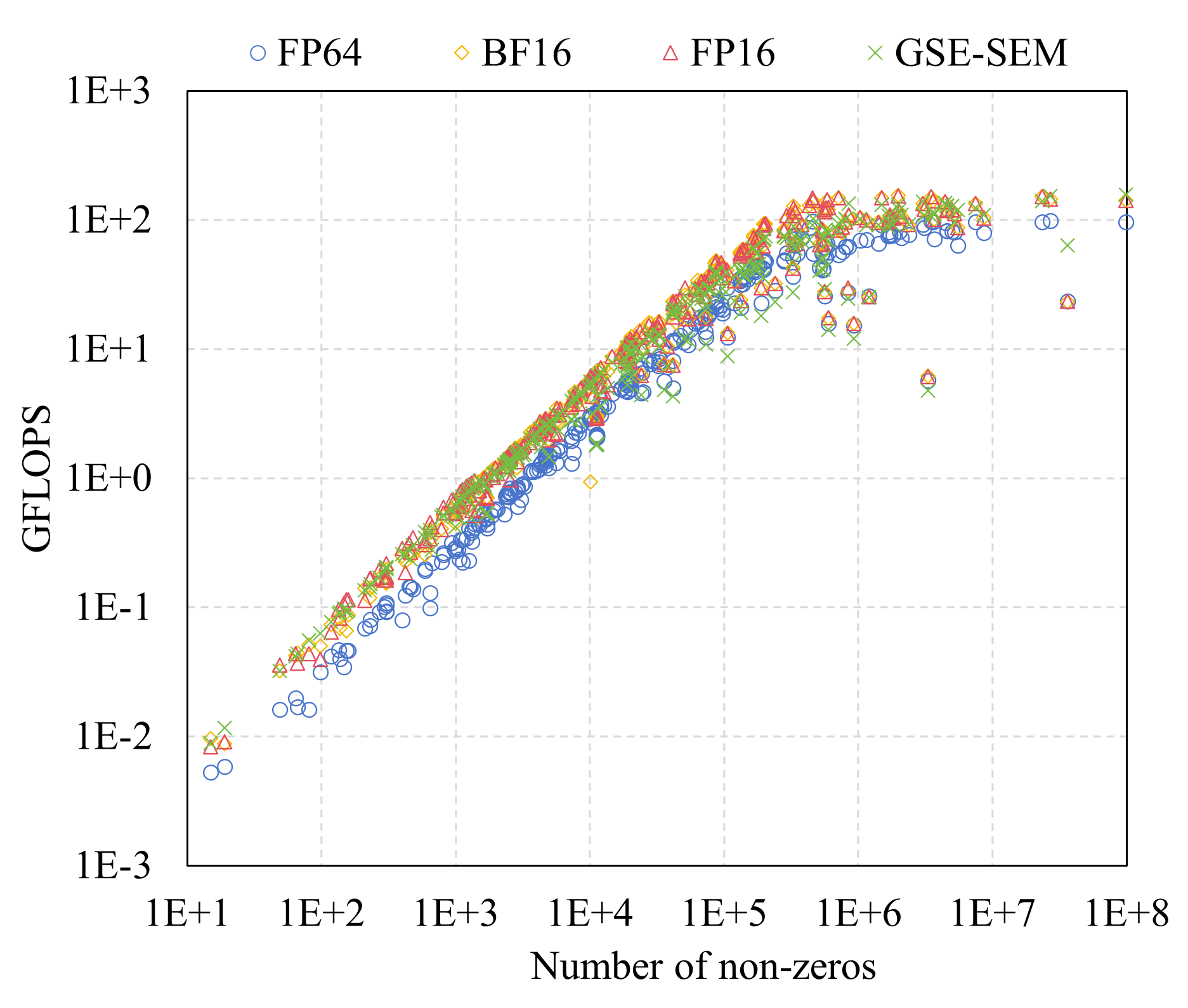}
    }
    \subfloat[Maximum absolute error]{
    \includegraphics[width=0.48\textwidth]{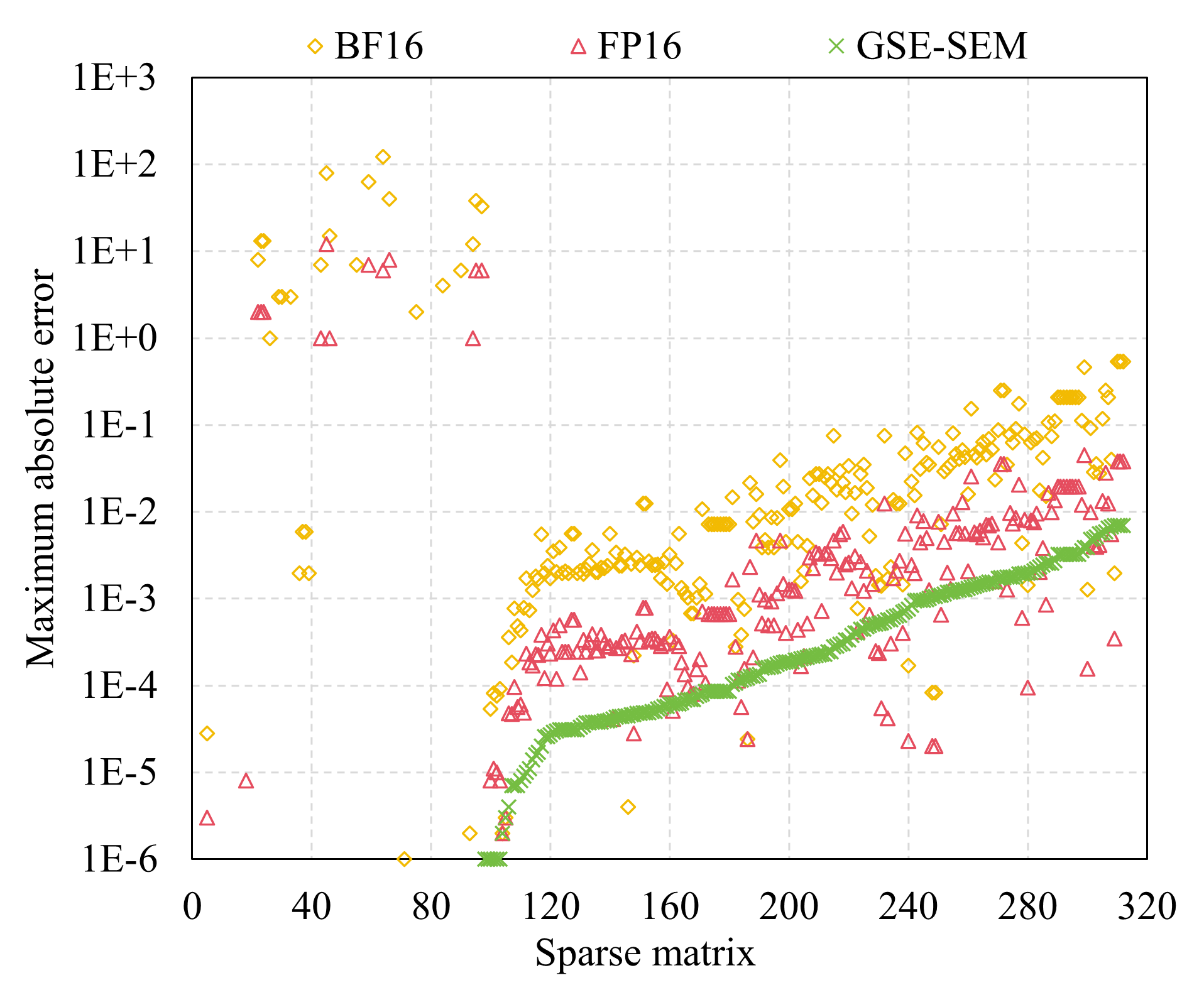}
    }
  \caption{Performance and error comparison of different SpMV algorithms on more than 300 sparse matrices.}
  \label{fig:spmv-comp}
\end{figure*}

It can be observed from Figure \ref{fig:shared_exponents}(a) that the performance difference between different $k$ is not significant. The average speedup first increases and then decreases as $k$ increases. On the one hand, as shown in Algorithm \ref{alg:spmv}, the floating-point numbers represented by GSE-SEM must be converted to FP64 representation on the GPU side before participating in the calculation. In the format conversion, we need to find the position where the first binary 1 appears from the most significant bit to the least significant bit in the head part of the GSE-SEM format. For floating-point numbers whose exponents are stored in the shared exponent array, the first binary 1 is at the most significant bit of the head part except for the sign bit, so the finding cost is relatively low. As $k$ increases, the exponents of more and more floating-point numbers are stored in the shared exponent array, decreasing the kernel execution cost and increasing the speedup.

On the other hand, as the number of shared exponents increases, the memory access cost of shared exponents gradually increases. The cost includes loading the shared exponents from global memory to shared memory and from shared memory to registers. Therefore, the speedup first increases and then decreases as $k$ increases. The average speedup is highest at $k$~=~8.

From Figure \ref{fig:shared_exponents}(b), it can be observed that the maximum absolute errors corresponding to $k$=2 and $k$=4 are usually larger than the errors of $k$=8, and the maximum absolute errors corresponding to $k$=16, $k$=32, and $k$=64 are smaller than the errors of $k$=8. From Figure \ref{fig:mean-spmv-speedup}, it can also be observed that as $k$ increases, the average error decreases. This is because exponent alignment and mantissa denormalization are required for floating-point numbers without corresponding exponents in the shared exponent array. Therefore, more shared exponents can reduce the loss of significant binary bits caused by exponent alignment and mantissa denormalization.

We set $k$ to 8 in subsequent experiments based on the above experimental results. Although the SpMV error at $k$=8 is larger than that at $k$=16, $k$=32, and $k$=64, we can utilize two tail segments of GSE-SEM to improve residuals in practical iterative applications.

\subsection{SpMV Performance Comparison}\label{sec:spmv-evaluation}

In this section, we compare the performance of different SpMV algorithms. In addition to FP64-SpMV, we also compare our method with two SpMV algorithms that use low-precision floating-point formats, FP16 and BF16, labeled as FP16-SpMV and BF16-SpMV, respectively. These two formats have the same bit-width as the head part of GSE-SEM. In the two SpMV algorithms, all non-zero elements are stored and loaded in FP16 or BF16 format, then converted to FP64 and multiplied by the double-precision vector. All intermediate results are accumulated in double-precision format to obtain the result vector. Figure \ref{fig:spmv-comp} compares the performance and errors of these SpMV algorithms for more than 300 sparse matrices. Figure \ref{fig:spmv-comp}(a) compares their GFLOPS with the results sorted in ascending order according to the number of non-zeros. Figure \ref{fig:spmv-comp}(b) presents the maximum absolute errors of the calculated vectors of BF16-SpMV, FP16-SpMV, and GSE-SEM-SpMV compared with FP64-SpMV, with the results sorted in ascending order according to GSE-SEM-SpMV.

We can observe that both FP16 and BF16 demonstrate comparable performance across all tested matrices. GSE-SEM only reads the head part and performs better than FP64-SpMV for most tested matrices. GSE-SEM-based SpMV is inferior to FP16-SpMV and BF16-SpMV in performance because they have almost the same memory access overhead. However, GSE-SEM-SpMV has a higher kernel execution overhead due to the need to convert floating-point numbers from GSE-SEM representation to FP64 representation. As presented in Figure \ref{fig:spmv-comp}(b), BF16-SpMV and FP16-SpMV exhibit larger errors than GSE-SEM-SpMV for most sparse matrices. For instance, the result vectors of GSE-SEM-SpMV are the same as those of FP64-SpMV for the first 97 sparse matrices, while the calculation errors of BF16-SpMV and FP16-SpMV on these matrices exceed 10 and even reach 100 on some matrices. In the following subsection, we apply these different SpMV algorithms in CG and GMRES solvers to demonstrate the advantages of the GSE-SEM representation.

\begin{figure*}
  \centering
  \subfloat[CG algorithm for the matrix \textit{consph}]{
    \includegraphics[width=0.49\textwidth]{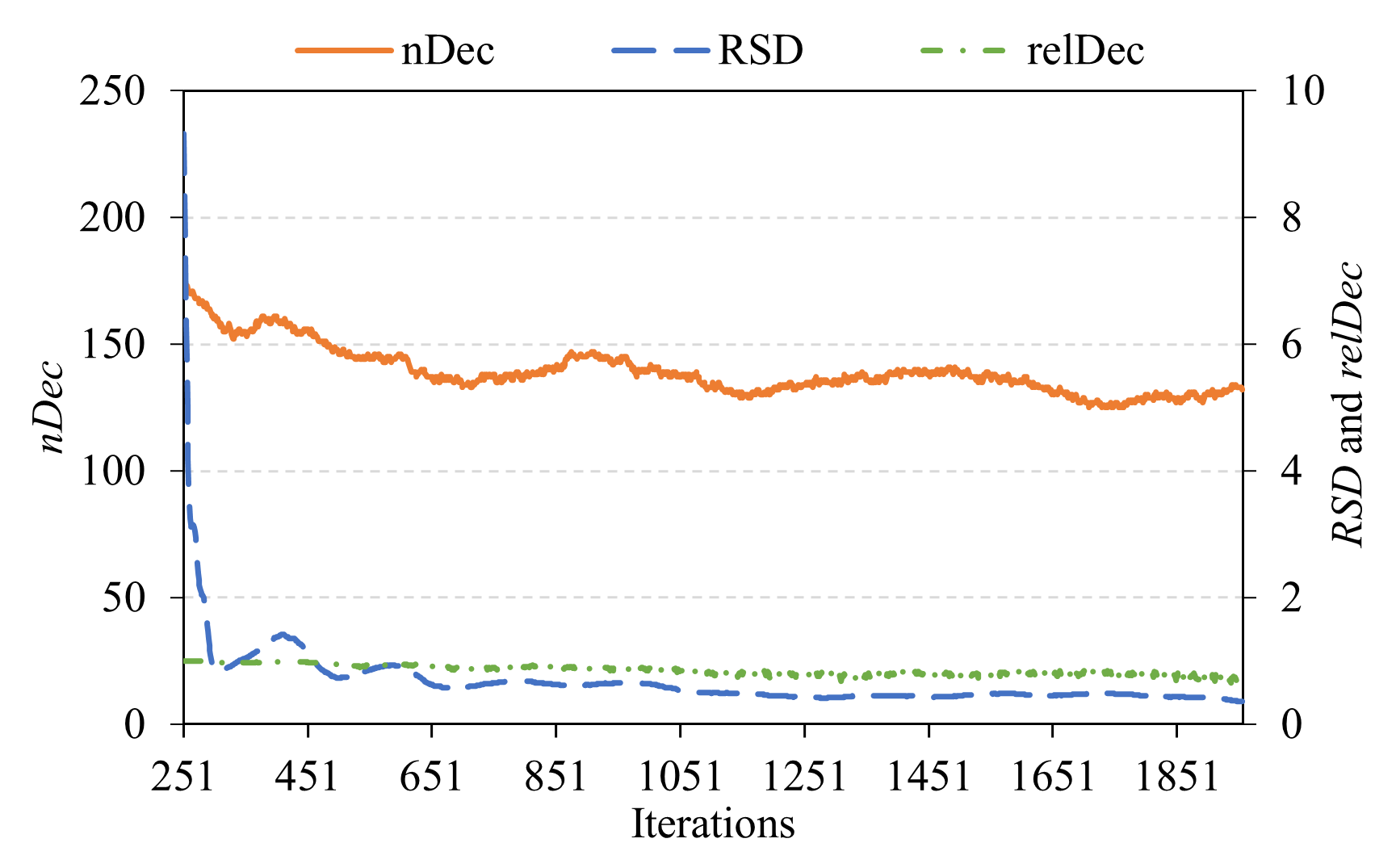}
    }
    \subfloat[CG algorithm for the matrix \textit{cvxbqp1}]{
    \includegraphics[width=0.49\textwidth]{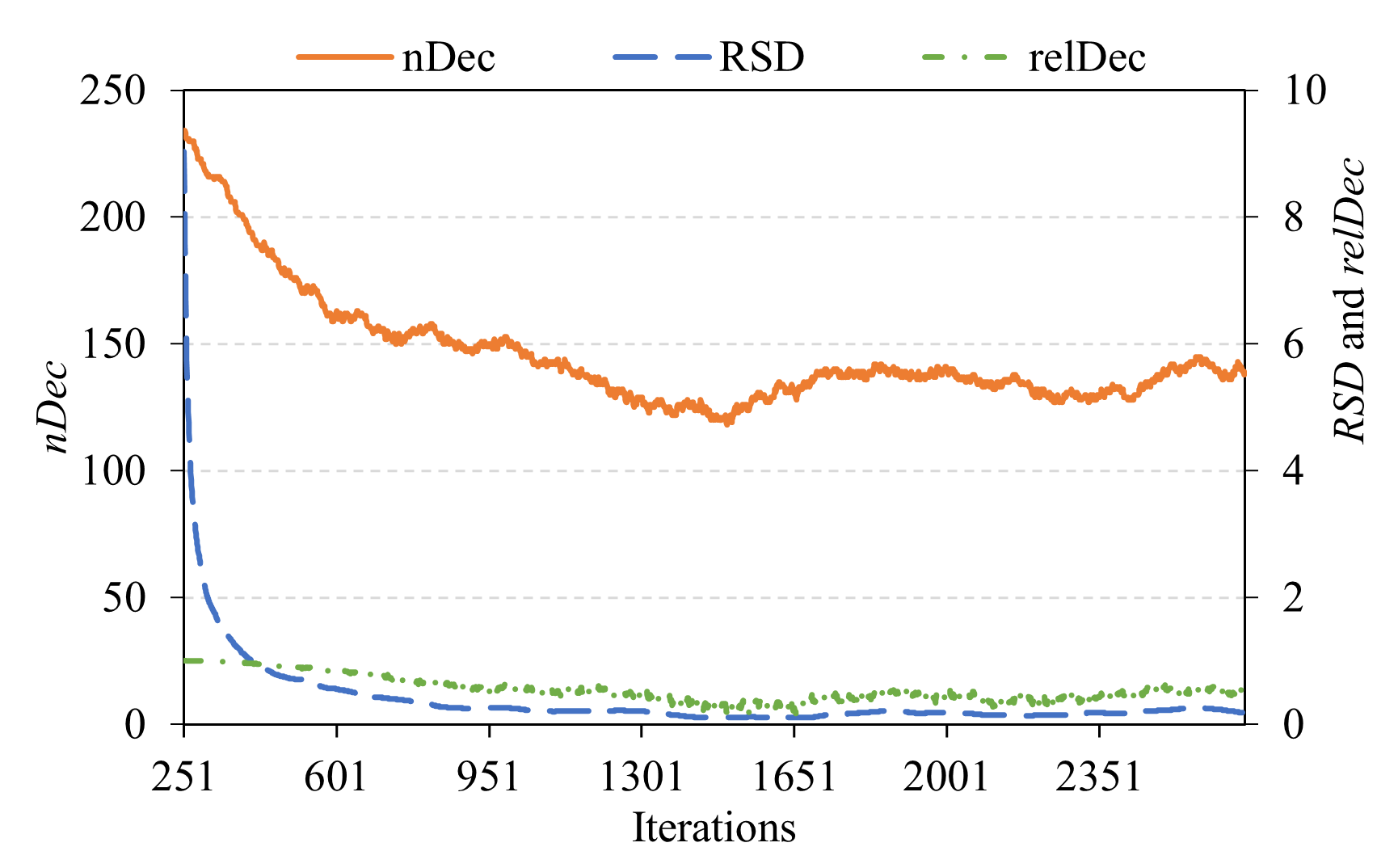}
    }
    \hfill
    \subfloat[GMRES algorithm for the matrix \textit{dw2048}]{
    \includegraphics[width=0.49\textwidth]{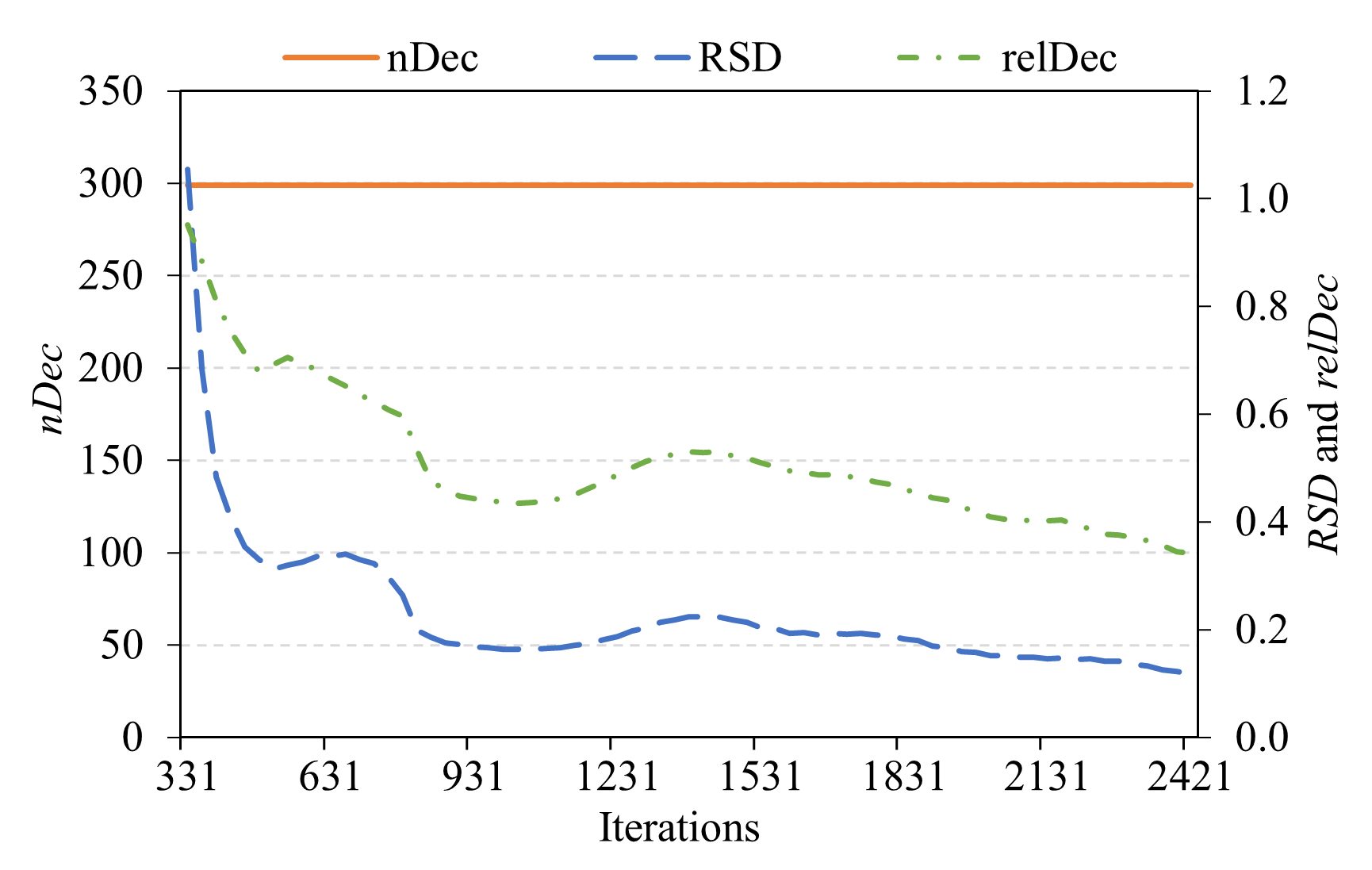}
    }
    \subfloat[GMRES algorithm for the matrix \textit{adder\_dcop\_01}]{
    \includegraphics[width=0.49\textwidth]{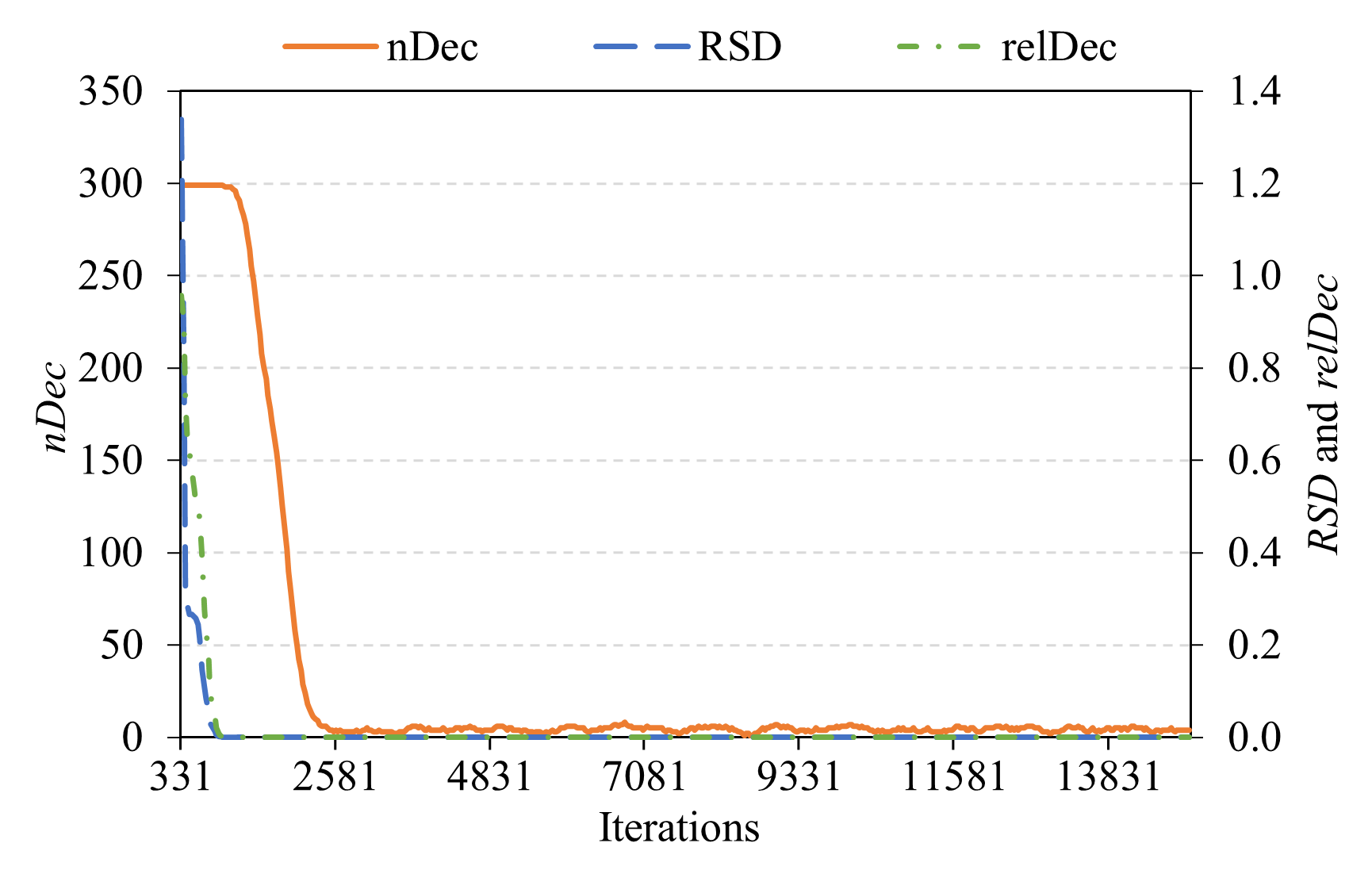}
    }
  \caption{Examples of parameter changes during the iterations of the CG and GMRES algorithms for solving different sparse linear systems.}
  \label{fig:params-example}
\end{figure*}

\subsection{Solvers Performance Comparison}\label{sec:solver-evaluation}

\subsubsection{Parameters Setting}
In the proposed stepped mixed-precision iterative algorithm, the precision switching mechanism includes six parameters: the initial low-precision iterations $l$, the historical iterations step $t$ for monitoring residuals, the iteration step $m$ for checking whether to increase precision, as well as the thresholds $RSD\_limit$, $nDec\_limit$, and $relDec\_limit$ that characterize the residual changes in the past $t$ iterations. The number of low-precision iterations $l$ cannot be too small; otherwise, the performance benefits of the stepped mixed-precision optimization are limited. In the experiments, we set $l$ to 9,000 and 3,000 in GMRES and CG, respectively. The number of historical iterations $t$ for residual monitoring cannot be too large; otherwise, it brings high computational overhead to record residual changes. In the experiment, $t$ is set to 300 and 250 in GMRES and CG, respectively. Additionally, $m$ is set to 1,500 and 500 in GMRES and CG, respectively.

As for the other three metrics, we collected the changes in $RSD$, $nDec$, and $relDec$ of GMRES and CG using FP64 representation based on the matrix sets presented in Table \ref{tab:matrix-list}. Four examples are presented in Figure \ref{fig:params-example}. In this figure, the upper two sub-figures present the experimental results of CG for matrices \textit{consph} and \textit{cvxbqp1}, and the lower two sub-figures present the experimental results of GMRES for matrices \textit{dw2048} and \textit{adder\_dcop\_01}. 

It can be observed that the change of the three parameters is different in the two iterative algorithms. As presented in Figures~\ref{fig:params-example}(a) and (b), in the CG algorithm, the number of decreases in the residual \textit{nDec} generally exhibits a declining trend, with intermittent fluctuations. For both example matrices, the final \textit{nDec} values are predominantly distributed within 100 to 150 iterations. The relative standard deviation of the residual \textit{RSD} is initially large and gradually decreases to a small value, indicating that the residual initially decreases rapidly. Similarly, the relative decrease rate of the residual \textit{relDec} shows a gradual decrease. In contrast, as presented in Figure~\ref{fig:params-example}(c), the number of decreases in the residual \textit{nDec} remains consistently at the historical iterations step of 300 for matrix \textit{dw2048} with GMRES, suggesting a continuous reduction in the residual throughout the iteration process. As presented in Figure \ref{fig:params-example}(d), the residual keeps decreasing in the historical 300 iterations at first and then only decreases a few times, so it can be observed that the \textit{RSD} in the later iterations is almost~0. Until the maximum iteration, the GMRES algorithm does not converge on this matrix.


In summary, the ranges and magnitudes for \textit{nDec}, \textit{RSD}, and \textit{relDec} in the two iterative algorithms are different. Consequently, different parameter settings are used for each algorithm to accommodate these differences. These three parameters are recorded in every $m$ iteration for each sparse matrix in the testing sets, and the average of each parameter is taken as the threshold in the subsequent experiments. Finally, for GMRES, $RSD\_limit$ is set to 0.03, $nDec\_limit$ is set to 80, and $relDec\_limit$ is set to 0.08. For CG, $RSD\_limit$ is set to 0.50, $nDec\_limit$ is set to 130, and $relDec\_limit$ is set to 0.45.

\subsubsection{Solver Convergence Comparison}

We compare the final relative residuals and the actual number of iterations for GMRES and CG algorithms using different floating-point formats, with the given error threshold and maximum iterations. Tables~\ref{tab:res-gmres} and \ref{tab:res-cg} provide the experimental results for GMRES and CG, respectively.


The tables show that FP16 leads to arithmetic overflow on 4 matrices in GMRES and 10 matrices in CG, while BF16 and GSE-SEM ran successfully on all tested matrices. Moreover, FP16 and BF16 achieved the smallest residual on 3 and 5 matrices, respectively, in GMRES, while GSE-SEM achieved the smallest residual on 7 matrices. In the CG solver, BF16 and GSE-SEM achieved the smallest residuals on 5 and 10 matrices, respectively, while FP16 did not demonstrate the smallest residual on any matrices. This indicates that our proposed GSE-SEM-based stepped mixed-precision iterative algorithm demonstrates superior numerical representation accuracy compared to the solvers based on the existing two 16-bit floating-point formats. Additionally, compared with FP64, GSE-SEM achieved convergence with fewer iterations on 10 matrices in GMRES and 2 matrices in CG, indicating that GSE-SEM can find faster convergence paths on some matrices. The phenomenon of a decrease in the number of iterations of the iterative algorithm after using mixed-precision optimization has also been observed in recent research~\cite{DBLP:journals/tpds/LindquistLD22,DBLP:journals/jip/ZhaoFI23}.

\begin{table}[!htbp]
\caption{Iterations and relative residuals of GMRES with different floating-point formats. Highlighted values in the \emph{Iterations} column represent that GSE-SEM takes fewer iterations to converge compared with FP64. Highlighted values in the \emph{Relative Residual} column represent the minimum relative residual among the FP16, BF16, and GSE-SEM.}
    \centering
    \scalebox{0.8}{
    \begin{tabular}{crrrr|rrrr}
    \toprule
        \multirow{2}{*}{ID} &
        \multicolumn{4}{c|}{Iterations} & \multicolumn{4}{c}{Relative Residual} \\ \cmidrule{2-9}
        & FP64 & FP16 & BF16 & GSE-SEM & FP64 & FP16 & BF16 & GSE-SEM \\ 
    \midrule
        1 & 2 & 2 & 2 & 2 & 4.2E-8 & \textbf{2.7E-7} & 3.8E-7 & 3.1E-7 \\ 
        2 & 2420 & 12195 & 9270 & \textbf{1657} & 1.0E-6 & 1.0E-6 & 1.0E-6 & \textbf{1.0E-6} \\ 
        3 & 2420 & 12195 & 9270 & \textbf{1657} & 1.0E-6 & 1.0E-6 & 1.0E-6 & \textbf{1.0E-6} \\ 
        4 & 15000 & 15000 & 15000 & \textbf{10640} & 1.3E-6 & 1.1E-6 & 1.3E-6 & \textbf{1.0E-6} \\ 
        5 & 15000 & 15000 & 15000 & \textbf{10640} & 1.3E-6 & 1.1E-6 & 1.3E-6 & \textbf{1.0E-6} \\ 
        6 & 1627 & 1027 & 7000 & \textbf{1148} & 1.0E-6 & 1.0E-6 & \textbf{1.0E-6} & 1.0E-6 \\ 
        7 & 438 & 15000 & 413 & \textbf{218} & 9.4E-7 & / & 1.0E-6 & \textbf{9.1E-7} \\ 
        8 & 55 & 55 & 56 & 55 & 9.1E-7 & 9.0E-7 & \textbf{7.5E-7} & 9.2E-7 \\ 
        9 & 5349 & 5542 & 15000 & \textbf{4522} & 1.0E-6 & \textbf{9.9E-7} & 3.1E-6 & 1.0E-6 \\ 
        10 & 247 & 249 & 387 & \textbf{246} & 9.9E-7 & \textbf{9.9E-7} & 9.9E-7 & 1.0E-6 \\ 
        11 & 58 & 76 & 81 & 58 & 9.9E-7 & 9.9E-7 & \textbf{9.1E-7} & 1.0E-6 \\ 
        12 & 55 & 15000 & 52 & \textbf{52} & 8.7E-7 & / & \textbf{7.5E-7} & 9.7E-7 \\ 
        13 & 500 & 1117 & 1743 & \textbf{384} & 1.0E-6 & 1.0E-6 & 1.0E-6 & \textbf{1.0E-6} \\ 
        14 & 12 & 15000 & 12 & 12 & 6.7E-7 & / & \textbf{6.6E-7} & 6.7E-7 \\ 
        15 & 518 & 15000 & 15000 & 531 & 1.0E-6 & / & 1.4E-6 & \textbf{1.0E-6} \\ 
        \bottomrule
    \end{tabular}}
    \label{tab:res-gmres}
\end{table}

\begin{table}[!htbp]
\caption{Iterations and relative residuals of CG with floating-point formats. Highlighted values in the \emph{Iterations} column represent that GSE-SEM takes fewer iterations to converge compared with FP64. Highlighted values in the \emph{Relative Residual} column represent the minimum relative residual among the FP16, BF16, and GSE-SEM.}
    \centering
    \scalebox{0.8}{
    \begin{tabular}{crrrr|rrrr}
    \toprule
        \multirow{2}{*}{ID} &
        \multicolumn{4}{c|}{Iterations} & \multicolumn{4}{c}{Relative Residual} \\ \cmidrule{2-9}
         & FP64 & FP16 & BF16 & GSE-SEM & FP64 & FP16 & BF16 & GSE-SEM \\
    \midrule 
        1 & 116 & 5000 & 104 & 116 & 8.2E-7 & / & \textbf{7.9E-7} & 8.2E-7 \\ 
        2 & 571 & 5000 & 513 & \textbf{509} & 6.9E-7 & / & 9.8E-7 & \textbf{9.3E-7} \\ 
        3 & 141 & 5000 & 139 & 144 & 9.4E-7 & / & 9.3E-7 & \textbf{5.9E-7} \\ 
        4 & 40 & 40 & 40 & 41 & 8.6E-7 & 8.7E-7 & 8.5E-7 & \textbf{8.1E-7} \\ 
        5 & 2684 & 5000 & 5000 & \textbf{2255} & 1.0E-6 & / & 3.5E-3 & \textbf{9.8E-7} \\ 
        6 & 1957 & 2198 & 5000 & 1986 & 1.0E-6 & 9.9E-7 & 2.0E-5 & \textbf{9.9E-7} \\ 
        7 & 5000 & 5000 & 5000 & 5000 & 4.2E-6 & / & \textbf{2.7E-2} & 6.0E-2 \\ 
        8 & 120 & 120 & 120 & 120 & 9.9E-7 & 9.9E-7 & \textbf{9.8E-7} & 9.9E-7 \\ 
        9 & 133 & 5000 & 5000 & 141 & 9.8E-7 & / & 4.4E-5 & \textbf{9.8E-7} \\ 
        10 & 129 & 5000 & 5000 & 136 & 9.7E-7 & / & 1.9E-5 & \textbf{9.6E-7} \\ 
        11 & 106 & 5000 & 5000 & 106 & 9.6E-7 & / & 1.3E-5 & \textbf{1.0E-6} \\ 
        12 & 5000 & 5000 & 5000 & 5000 & 8.7E-6 & / & \textbf{4.5E-3} & 3.9E-1 \\ 
        13 & 180 & 332 & 5000 & 187 & 9.9E-7 & 1.0E-6 & 1.3E-3 & \textbf{1.0E-6} \\ 
        14 & 228 & 3042 & 5000 & 230 & 1.0E-6 & 1.0E-6 & 1.4E-5 & \textbf{1.0E-6} \\ 
        15 & 5000 & 5000 & 5000 & 5000 & 1.3E-4 & / & \textbf{4.3E-2} & 5.4E-1 \\
        \bottomrule
    \end{tabular}}
    \label{tab:res-cg}
\end{table}

\subsubsection{Solver Time Comparison}
In this subsection, we compare the running time of GMRES and CG. Except for the SpMV operator, all other vector operations are performed using the FP64 format. In SpMV, sparse matrices of different precisions are read from memory, including FP64, FP16, BF16, and GSE-SEM, and then their non-zeros are converted to FP64 and used in the calculation.


\begin{figure}[!htbp]
    \centering
    \includegraphics[width=0.99\linewidth]{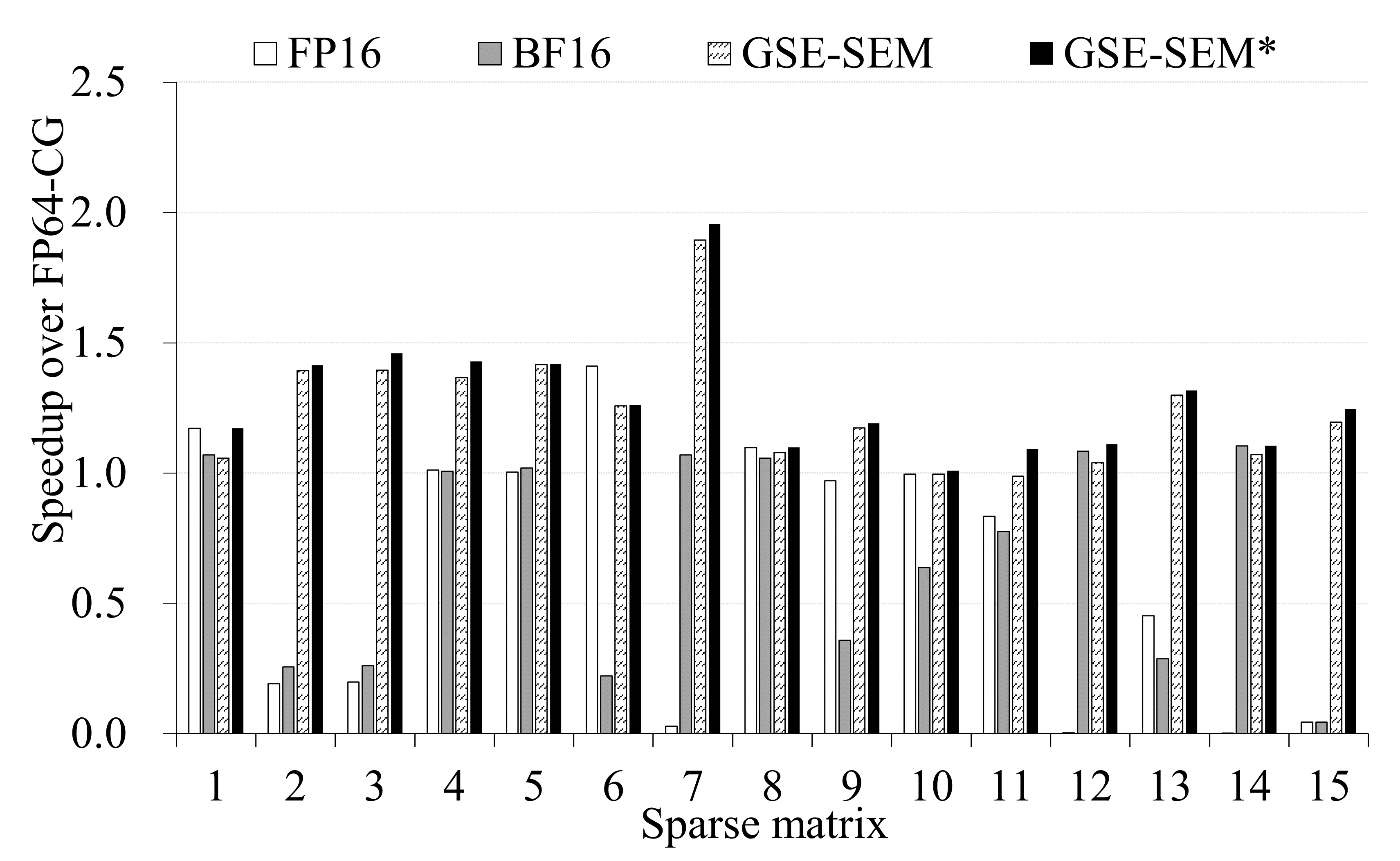}
    \caption{Performance comparison of GMRES with SpMV using different floating-point representations. GSE-SEM*-based GMRES is similar to GMRES based on GSE-SEM, but does not consider the overhead of format conversion in the SpMV kernel.}
    \label{fig:gmres-speedup}
\end{figure}

Figure~\ref{fig:gmres-speedup} presents the experimental results of GMRES. We can observe that the performance of FP16 is inferior to that of FP64 on 8 matrices (matrices 2, 3, 7, 11, 12, 13, 14, and 15) due to the increased number of iterations. For the same reason, BF16 achieves a speedup of less than 1 on 8 matrices (matrices 2, 3, 6, 9, 10, 11, 13, and 15). In contrast, GSE-SEM demonstrates the best overall performance and achieves an average speedup of 1.24x on the testing set of GMRES, while the average speedups of FP16 and BF16 are 0.61x and 0.67x, respectively. Notably, for the seventh matrix, GSE-SEM achieves the highest speedup of 1.89x because FP64 requires 438 iterations to achieve convergence, while GSE-SEM only requires 218 iterations to achieve a similar residual.

\begin{figure}[!htbp]
    \centering
    \includegraphics[width=0.99\linewidth]{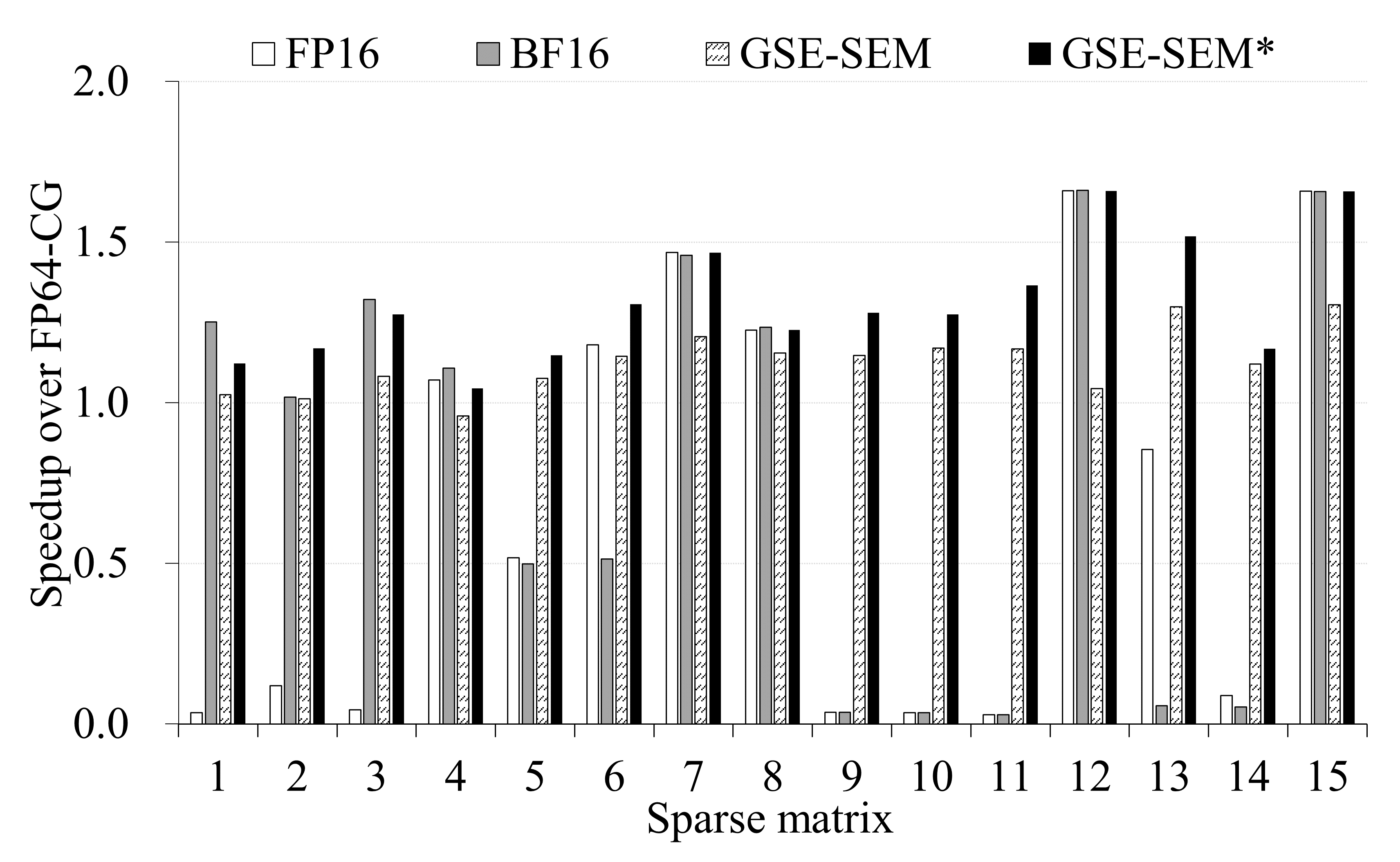}
    \caption{Performance comparison of CG with SpMV using different floating-point representations. GSE-SEM*-based CG is similar to CG based on GSE-SEM but does not consider the overhead of format conversion in the SpMV kernel.}
    \label{fig:cg-speedup}
\end{figure}

Figure \ref{fig:cg-speedup} presents the experimental results of CG. Similar to GMRES, due to the increased iterations, FP16 and BF16 achieve a speedup of less than 1 on about 9 matrices (matrices 1, 2, 3, 5, 9, 10, 11, 13, and 14) and 7 matrices (matrices 5, 6, 9, 10, 11, 13, and 14), respectively. The speedup of GSE-SEM is greater than 1 on most sparse matrices. Overall, the average speedups of FP16, BF16, and GSE-SEM over FP64 are 0.66x, 0.76x, and 1.13x, respectively.

We currently implement the GSE-SEM format using software simulation, so part of the performance benefits is offset by the format conversion overhead in the SpMV kernel. We calculated the performance improvement brought by the GSE-SEM format and stepped mixed-precision optimization when the format conversion overhead is not considered, represented by GSE-SEM* in Figures \ref{fig:gmres-speedup} and \ref{fig:cg-speedup}. The time corresponding to GSE-SEM* is calculated by Equation \ref{equ:GSE-SEM*}, where $TIME_{FP16}$ represents the time of GMRES or CG based on FP16, and $ITERS_{FP16}$ and $ITERS_{GSE-SEM}$ represent the iterations of GMRES or CG based on FP16 and GSE-SEM, respectively. As expected, we can observe that GSE-SEM* achieves higher speedups than GSE-SEM on both solvers. Unlike GMRES, SpMV usually takes a higher proportion of time in the CG algorithm. Therefore, more performance gains can be observed in the CG algorithm when the format conversion overhead is not considered. The average speedups of GSE-SEM* on GMRES and CG are 1.29x and 1.31x, respectively. Taking matrix 13 in the CG algorithm as an example, if we do not consider the format conversion overhead in the GSE-SEM-based SpMV, the proposed GMRES can achieve a speedup of 1.52x, higher than the original 1.30x. Therefore, if the GSE-SEM format proposed in this paper can be supported by hardware, we can observe more performance gains on these popular solvers.
\begin{equation}\label{equ:GSE-SEM*}
    \frac{TIME_{FP16}}{ITERS_{FP16}}\times ITERS_{GSE-SEM}
\end{equation}

\section{Related Works}\label{sec:related-work}
On the one hand, the calculations in iterative algorithms are usually floating-point operations, so the selection of floating-point formats plays a crucial role in their performance. On the other hand, SpMV, one of the most important computational kernels in iterative algorithms, has received intensive attention for mixed-precision optimization. Therefore, we introduce the related work from three perspectives: new floating-point format design, mixed-precision SpMV optimization, and mixed-precision iterative solver optimization.

\subsection{New Floating-Point Format Design}
Grützmacher et al.~\cite{DBLP:journals/concurrency/GrutzmacherCFGA20} segment the FP64 representation into two 32-bit segments, the head and tail. When the application requires lower precision, only the head segment is read. In cases requiring higher precision, both head and tail segments are accessed separately and concatenated. The work by Agrawal et al.~\cite{agrawal2019dlfloat} proposes a 16-bit floating-point format for deep learning applications. This format consists of 6 exponent bits and 9 mantissa bits. Sadi et al.~\cite{hassani2024novel} introduce an 8-bit floating-point (FP8) format, which dynamically adjusts the bias value of FP8 to accommodate the numerical representation range required by DNN parameters and input feature maps. Hajizadeh et al.~\cite{hajizadeh2024cufp} propose a customizable floating-point library for HLS in FPGA applications, allowing programmers to define exponent and mantissa bit-widths at compile time. Optimized implementations of key operations like vector summation and matrix-vector multiplication present reduced latency and resource usage compared to vendor IP blocks, enhancing FPGA computational efficiency. Tan et al.~\cite{tan2024low} introduce a novel low-cost floating-point FMA (fused multiply-add) unit supporting single-precision, TF32, BF16, half-precision, and double-precision formats, enhancing performance through pipelining and parallel processing. It integrates iteration and vectorization techniques for optimized HPC-AI applications, achieving up to 1.5x performance gains over traditional designs.

\subsection{Mixed-Precision SpMV}
The MpSpMV algorithm, proposed by Ahmad et al.~\cite{10.1145/3371275}, uses single and double precision for elements within and outside a specified range, respectively. The row-wise partitioned SpMV, proposed by Tezcan et al.~\cite{DBLP:conf/sbac-pad/TezcanTKKU22}, uses single or double-precision for matrix rows based on their non-zeros' distribution. Further, Graillat et al.~\cite{graillat:hal-03561193} employ a precision-partitioning strategy for non-zero elements to optimize storage and computation. They also introduce adaptive precision algorithms for SpMV in another work \cite{graillat2024reduced}, integrating custom reduced precision formats and optimized memory access techniques. They explore diverse significand and exponent bit allocations to enhance performance, surpassing traditional double and single precision in computational efficiency for various applications. Zhao et al. \cite{zhao2024block} propose the BDMP method, which segments matrices into blocks with tailored precision, alongside two computation algorithms to improve parallelism and memory efficiency. Mukunoki et al.~\cite{mukunoki2023sparse} also contribute by examining the benefits of reduced-precision formats in SpMV, highlighting the potential for computational savings. Investigations into precision variability by Isupov~\cite{DBLP:journals/jocs/Isupov22} and Kouya~\cite{DBLP:journals/corr/Kouya14a} explore the efficacy of different bit-widths and precision levels in SpMV algorithms. Isupov~\cite{DBLP:journals/jocs/Isupov22} tests various precision bit-widths ranging from 106 to 848 in different SpMV algorithms by software simulations. The sparse matrix is consistently represented in double precision, and the multiplication and result vectors are represented in different high-precision formats.

\subsection{Mixed-Precision Iterative Algorithms}
In iterative algorithms, leveraging mixed-precision computing has shown promise for enhancing computational efficiency and maintaining accuracy. Key developments include the GMRES-IR by Carson et al.~\cite{DBLP:journals/siamsc/CarsonH18}, which utilizes different floating-point precision for various stages of computation, and further optimizations for GMRES that introduce lower precision in specific operations for efficiency gains. They also explore a five-precision iterative refinement scheme for solving sparse linear systems using sparse approximate inverse preconditioners \cite{carson2023mixed}. This study investigates computation constraints and convergence behavior, highlighting trade-offs between preconditioner sparsity and GMRES performance. Lindquist et al.~\cite{lindquist2020improving} and subsequent studies \cite{DBLP:journals/tpds/LindquistLD22} have showcased mixed-precision GMRES implementations focusing on precision placement to balance performance and accuracy. Furthermore, the architectural benefits of mixed-precision arithmetic, as discussed by Baboulin et al.~\cite{DBLP:journals/cphysics/BaboulinBDKLLLT09}, emphasize the speed advantages on modern hardware while maintaining result integrity. Amestoy et al.~\cite{amestoy2023combining} propose a framework for parallel numerical algorithms that enhance accuracy and computational speed through error-free transformation and mixed precision. Aliaga et al.~\cite{aliaga2023compressed} present a novel compressed GMRES solver utilizing Ginkgo's~\cite{ginkgo-toms-2022} memory accessor to enhance the efficiency of solving large-scale sparse linear systems. By decoupling the storage format of the orthogonal basis from the arithmetic precision used in operations, the solver effectively reduces communication and memory access costs. Zhao et al.~\cite{DBLP:journals/jip/ZhaoFI23} focus on mixed-precision iterative refinement methods for solving large sparse linear systems, utilizing BiCGSTAB with FP32 as an inner solver.


In summary, most mixed-precision optimization methods are based on existing general floating-point formats and do not fully utilize the numerical distribution features of specific applications.

\section{Conclusion}\label{sec:conclusion}
Given the redundancy in the exponent part of existing IEEE 754 floating-point formats, this paper introduces an innovative group-shared exponent representation. By identifying the most frequent exponents in a set of floating-point numbers as shared exponents and utilizing denormalized floating-point representation, we effectively enable shared exponents to represent all non-zeros. Additionally, we combine this format with a mantissa segmentation scheme to meet diverse precision requirements across different applications, eliminating redundant data storage. Our proposed stepped mixed-precision optimization method, based on the new floating-point format, enhances iterative algorithms. Performance comparisons on CPU+GPU demonstrate that the group-shared floating-point format achieves not only superior convergence in GMRES compared to BF16 and FP16 formats but also a remarkable improvement in performance over FP64, underscoring its potential to redefine efficiency and accuracy in high-performance computing applications.

\bibliographystyle{IEEEtran}
\bibliography{IEEEabrv,sample-base}

\begin{thebibliography}{10}
\providecommand{\url}[1]{#1}
\csname url@samestyle\endcsname
\providecommand{\newblock}{\relax}
\providecommand{\bibinfo}[2]{#2}
\providecommand{\BIBentrySTDinterwordspacing}{\spaceskip=0pt\relax}
\providecommand{\BIBentryALTinterwordstretchfactor}{4}
\providecommand{\BIBentryALTinterwordspacing}{\spaceskip=\fontdimen2\font plus
\BIBentryALTinterwordstretchfactor\fontdimen3\font minus \fontdimen4\font\relax}
\providecommand{\BIBforeignlanguage}[2]{{%
\expandafter\ifx\csname l@#1\endcsname\relax
\typeout{** WARNING: IEEEtran.bst: No hyphenation pattern has been}%
\typeout{** loaded for the language `#1'. Using the pattern for}%
\typeout{** the default language instead.}%
\else
\language=\csname l@#1\endcsname
\fi
#2}}
\providecommand{\BIBdecl}{\relax}
\BIBdecl

\bibitem{10.1145/3371275}
\BIBentryALTinterwordspacing
K.~Ahmad, H.~Sundar, and M.~Hall, ``Data-driven mixed precision sparse matrix vector multiplication for gpus,'' \emph{ACM Trans. Archit. Code Optim.}, vol.~16, no.~4, dec 2019. [Online]. Available: \url{https://doi.org/10.1145/3371275}
\BIBentrySTDinterwordspacing

\bibitem{DBLP:journals/jcam/MukunokiO20}
\BIBentryALTinterwordspacing
D.~Mukunoki and T.~Ogita, ``Performance and energy consumption of accurate and mixed-precision linear algebra kernels on gpus,'' \emph{J. Comput. Appl. Math.}, vol. 372, p. 112701, 2020. [Online]. Available: \url{https://doi.org/10.1016/j.cam.2019.112701}
\BIBentrySTDinterwordspacing

\bibitem{DBLP:conf/sbac-pad/TezcanTKKU22}
\BIBentryALTinterwordspacing
E.~Tezcan, T.~Torun, F.~Kosar, K.~Kaya, and D.~Unat, ``Mixed and multi-precision spmv for gpus with row-wise precision selection,'' in \emph{2022 {IEEE} 34th International Symposium on Computer Architecture and High Performance Computing (SBAC-PAD), Bordeaux, France, November 2-5, 2022}.\hskip 1em plus 0.5em minus 0.4em\relax {IEEE}, 2022, pp. 31--40. [Online]. Available: \url{https://doi.org/10.1109/SBAC-PAD55451.2022.00014}
\BIBentrySTDinterwordspacing

\bibitem{graillat:hal-03561193}
\BIBentryALTinterwordspacing
S.~Graillat, F.~J{\'e}z{\'e}quel, T.~Mary, and R.~Molina, ``{Adaptive precision matrix-vector product},'' Feb. 2022, working paper or preprint. [Online]. Available: \url{https://hal.science/hal-03561193}
\BIBentrySTDinterwordspacing

\bibitem{DBLP:journals/jocs/Isupov22}
\BIBentryALTinterwordspacing
K.~Isupov, ``Multiple-precision sparse matrix-vector multiplication on gpus,'' \emph{J. Comput. Sci.}, vol.~61, p. 101609, 2022. [Online]. Available: \url{https://doi.org/10.1016/j.jocs.2022.101609}
\BIBentrySTDinterwordspacing

\bibitem{DBLP:journals/corr/Kouya14a}
\BIBentryALTinterwordspacing
T.~Kouya, ``A highly efficient implementation of multiple precision sparse matrix-vector multiplication and its application to product-type krylov subspace methods,'' \emph{CoRR}, vol. abs/1411.2377, 2014. [Online]. Available: \url{http://arxiv.org/abs/1411.2377}
\BIBentrySTDinterwordspacing

\bibitem{DBLP:conf/cluster/MukunokiI16}
\BIBentryALTinterwordspacing
D.~Mukunoki and T.~Imamura, ``Reduced-precision floating-point formats on gpus for high performance and energy efficient computation,'' in \emph{2016 {IEEE} International Conference on Cluster Computing, {CLUSTER} 2016, Taipei, Taiwan, September 12-16, 2016}.\hskip 1em plus 0.5em minus 0.4em\relax {IEEE} Computer Society, 2016, pp. 144--145. [Online]. Available: \url{https://doi.org/10.1109/CLUSTER.2016.77}
\BIBentrySTDinterwordspacing

\bibitem{graillat2024reduced}
S.~Graillat, F.~J{\'e}z{\'e}quel, T.~Mary, R.~Molina, and D.~Mukunoki, ``Reduced-precision and reduced-exponent formats for accelerating adaptive precision sparse matrix-vector product,'' 2024.

\bibitem{zhao2024block}
Z.~Zhao, G.~Zhang, Y.~Wu, R.~Hong, Y.~Yang, and Y.~Fu, ``Block-wise dynamic mixed-precision for sparse matrix-vector multiplication on gpus,'' \emph{The Journal of Supercomputing}, pp. 1--33, 2024.

\bibitem{mukunoki2023sparse}
D.~Mukunoki, M.~Kawai, and T.~Imamura, ``Sparse matrix-vector multiplication with reduced-precision memory accessor,'' in \emph{2023 IEEE 16th International Symposium on Embedded Multicore/Many-core Systems-on-Chip (MCSoC)}.\hskip 1em plus 0.5em minus 0.4em\relax IEEE, 2023, pp. 608--615.

\bibitem{DBLP:journals/siamsc/CarsonH18}
\BIBentryALTinterwordspacing
E.~C. Carson and N.~J. Higham, ``Accelerating the solution of linear systems by iterative refinement in three precisions,'' \emph{{SIAM} J. Sci. Comput.}, vol.~40, no.~2, 2018. [Online]. Available: \url{https://doi.org/10.1137/17M1140819}
\BIBentrySTDinterwordspacing

\bibitem{DBLP:journals/corr/abs-1907-10550}
\BIBentryALTinterwordspacing
S.~Gratton, E.~Simon, D.~Titley{-}P{\'{e}}loquin, and P.~L. Toint, ``Exploiting variable precision in {GMRES},'' \emph{CoRR}, vol. abs/1907.10550, 2019. [Online]. Available: \url{http://arxiv.org/abs/1907.10550}
\BIBentrySTDinterwordspacing

\bibitem{DBLP:journals/corr/abs-2009-12101}
\BIBentryALTinterwordspacing
J.~I. Aliaga, H.~Anzt, T.~Gr{\"{u}}tzmacher, E.~S. Quintana{-}Ort{\'{\i}}, and A.~E. Tom{\'{a}}s, ``Compressed basis {GMRES} on high performance gpus,'' \emph{CoRR}, vol. abs/2009.12101, 2020. [Online]. Available: \url{https://arxiv.org/abs/2009.12101}
\BIBentrySTDinterwordspacing

\bibitem{le2018mixed}
M.~Le~Gallo, A.~Sebastian, R.~Mathis, M.~Manica, H.~Giefers, T.~Tuma, C.~Bekas, A.~Curioni, and E.~Eleftheriou, ``Mixed-precision in-memory computing,'' \emph{Nature Electronics}, vol.~1, no.~4, pp. 246--253, 2018.

\bibitem{loe2021study}
J.~A. Loe, C.~A. Glusa, I.~Yamazaki, E.~G. Boman, and S.~Rajamanickam, ``A study of mixed precision strategies for gmres on gpus,'' \emph{arXiv preprint arXiv:2109.01232}, 2021.

\bibitem{Davis2011_SuiteSparse}
\BIBentryALTinterwordspacing
T.~A. Davis and Y.~Hu, ``{The University of Florida Sparse Matrix Collection},'' \emph{{ACM} Trans. Math. Software}, vol.~38, no.~1, pp. 1--25, 2011. [Online]. Available: \url{https://doi.org/10.1145/2049662.2049663}
\BIBentrySTDinterwordspacing

\bibitem{DBLP:journals/concurrency/GrutzmacherCFGA20}
\BIBentryALTinterwordspacing
T.~Gr{\"{u}}tzmacher, T.~Cojean, G.~Flegar, F.~G{\"{o}}bel, and H.~Anzt, ``A customized precision format based on mantissa segmentation for accelerating sparse linear algebra,'' \emph{Concurr. Comput. Pract. Exp.}, vol.~32, no.~15, 2020. [Online]. Available: \url{https://doi.org/10.1002/cpe.5418}
\BIBentrySTDinterwordspacing

\bibitem{Cusp}
\BIBentryALTinterwordspacing
S.~Dalton, N.~Bell, L.~Olson, and M.~Garland, ``{Cusp: Generic Parallel Algorithms for Sparse Matrix and Graph Computations},'' 2014, version 0.5.0. [Online]. Available: \url{http://cusplibrary.github.io/}
\BIBentrySTDinterwordspacing

\bibitem{DBLP:conf/icpads/GaoJLSWS21}
\BIBentryALTinterwordspacing
J.~Gao, W.~Ji, J.~Liu, S.~Shao, Y.~Wang, and F.~Shi, ``{AMF-CSR:} adaptive multi-row folding of {CSR} for spmv on {GPU},'' in \emph{27th {IEEE} International Conference on Parallel and Distributed Systems, {ICPADS} 2021, Beijing, China, December 14-16, 2021}.\hskip 1em plus 0.5em minus 0.4em\relax {IEEE}, 2021, pp. 418--425. [Online]. Available: \url{https://doi.org/10.1109/ICPADS53394.2021.00058}
\BIBentrySTDinterwordspacing

\bibitem{DBLP:journals/tpds/LindquistLD22}
\BIBentryALTinterwordspacing
N.~Lindquist, P.~Luszczek, and J.~J. Dongarra, ``Accelerating restarted {GMRES} with mixed precision arithmetic,'' \emph{{IEEE} Trans. Parallel Distributed Syst.}, vol.~33, no.~4, pp. 1027--1037, 2022. [Online]. Available: \url{https://doi.org/10.1109/TPDS.2021.3090757}
\BIBentrySTDinterwordspacing

\bibitem{DBLP:journals/jip/ZhaoFI23}
\BIBentryALTinterwordspacing
Y.~Zhao, T.~Fukaya, and T.~Iwashita, ``Numerical behavior of mixed precision iterative refinement using the bicgstab method,'' \emph{J. Inf. Process.}, vol.~31, pp. 860--874, 2023. [Online]. Available: \url{https://doi.org/10.2197/ipsjjip.31.860}
\BIBentrySTDinterwordspacing

\bibitem{agrawal2019dlfloat}
A.~Agrawal, S.~M. Mueller, B.~M. Fleischer, X.~Sun, N.~Wang, J.~Choi, and K.~Gopalakrishnan, ``Dlfloat: A 16-b floating point format designed for deep learning training and inference,'' in \emph{2019 IEEE 26th Symposium on Computer Arithmetic (ARITH)}.\hskip 1em plus 0.5em minus 0.4em\relax IEEE, 2019, pp. 92--95.

\bibitem{hassani2024novel}
M.~Hassani~Sadi, C.~Sudarshan, and N.~Wehn, ``Novel adaptive quantization methodology for 8-bit floating-point dnn training,'' \emph{Design Automation for Embedded Systems}, pp. 1--20, 2024.

\bibitem{hajizadeh2024cufp}
\BIBentryALTinterwordspacing
F.~Hajizadeh, T.~Ould-Bachir, and J.-P. David, ``Cufp: An hls library for customized floating-point operators,'' \emph{Preprints}, June 2024. [Online]. Available: \url{https://doi.org/10.20944/preprints202406.1239.v1}
\BIBentrySTDinterwordspacing

\bibitem{tan2024low}
H.~Tan, J.~Zhang, L.~Huang, X.~He, Y.~Wang, and L.~Xiao, ``A low-cost floating-point fma unit supporting package operations for hpc-ai applications,'' \emph{IEEE Transactions on Circuits and Systems II: Express Briefs}, pp. 1--1, 2024.

\bibitem{carson2023mixed}
\BIBentryALTinterwordspacing
E.~Carson and N.~Khan, ``Mixed precision iterative refinement with sparse approximate inverse preconditioning,'' \emph{SIAM Journal on Scientific Computing}, vol.~45, no.~3, pp. C131--C153, 2023. [Online]. Available: \url{https://doi.org/10.1137/22M1487709}
\BIBentrySTDinterwordspacing

\bibitem{lindquist2020improving}
N.~Lindquist, P.~Luszczek, and J.~Dongarra, ``Improving the performance of the gmres method using mixed-precision techniques,'' in \emph{Driving Scientific and Engineering Discoveries Through the Convergence of HPC, Big Data and AI: 17th Smoky Mountains Computational Sciences and Engineering Conference, SMC 2020, Oak Ridge, TN, USA, August 26-28, 2020, Revised Selected Papers 17}.\hskip 1em plus 0.5em minus 0.4em\relax Springer, 2020, pp. 51--66.

\bibitem{DBLP:journals/cphysics/BaboulinBDKLLLT09}
\BIBentryALTinterwordspacing
M.~Baboulin, A.~Buttari, J.~J. Dongarra, J.~Kurzak, J.~Langou, J.~Langou, P.~Luszczek, and S.~Tomov, ``Accelerating scientific computations with mixed precision algorithms,'' \emph{Comput. Phys. Commun.}, vol. 180, no.~12, pp. 2526--2533, 2009. [Online]. Available: \url{https://doi.org/10.1016/j.cpc.2008.11.005}
\BIBentrySTDinterwordspacing

\bibitem{amestoy2023combining}
P.~Amestoy, A.~Buttari, N.~J. Higham, J.-Y. L’excellent, T.~Mary, and B.~Vieuble, ``Combining sparse approximate factorizations with mixed-precision iterative refinement,'' \emph{ACM Transactions on Mathematical Software}, vol.~49, no.~1, pp. 1--29, 2023.

\bibitem{aliaga2023compressed}
J.~I. Aliaga, H.~Anzt, T.~Gr{\"u}tzmacher, E.~S. Quintana-Ort{\'\i}, and A.~E. Tom{\'a}s, ``Compressed basis gmres on high-performance graphics processing units,'' \emph{The International Journal of High Performance Computing Applications}, vol.~37, no.~2, pp. 82--100, 2023.

\bibitem{ginkgo-toms-2022}
\BIBentryALTinterwordspacing
H.~Anzt, T.~Cojean, G.~Flegar, F.~Göbel, T.~Grützmacher, P.~Nayak, T.~Ribizel, Y.~M. Tsai, and E.~S. Quintana-Ortí, ``{Ginkgo: A Modern Linear Operator Algebra Framework for High Performance Computing},'' \emph{ACM Transactions on Mathematical Software}, vol.~48, no.~1, pp. 2:1--2:33, Feb. 2022. [Online]. Available: \url{https://doi.org/10.1145/3480935}
\BIBentrySTDinterwordspacing

\end{thebibliography}

\end{document}